\documentclass[%
reprint,
nofootinbib,
amsmath,amssymb,
aps,
]{revtex4-2}
\usepackage[colorlinks=true,linkcolor=blue,citecolor=blue,urlcolor=
blue]{hyperref}

\usepackage{graphicx}
\usepackage{dcolumn}
\usepackage{bm}
\usepackage{amsmath}
\usepackage[dvipsnames]{xcolor}
\usepackage{subcaption}
\usepackage{lipsum}
\usepackage{mwe}

\begin{document}
	
	\preprint{APS/123-QED}
	
	\title{Wave Phenomena In General Relativistic Magnetohydrodynamics}
	
	\author{Ankit Kumar Panda}
	\email{ankitkumar.panda@niser.ac.in}
	\author{Victor Roy}%
	\email{victor@niser.ac.in}
	\affiliation{School of Physical Sciences, National Institute of Science Education and Research,
		An OCC of Homi Bhabha National Institute, Jatni-752050, India}%

	\date{\today}
	
	\begin{abstract}
		Here we study the wave propagation and stability of general relativistic non-resistive dissipative second-order magnetohydrodynamic equations in curved space-time. We solve the Boltzmann equation for a system of particles and antiparticles using the relaxation time approximation and the Chapman-Enskog-like gradient expansion for the off-equilibrium distribution function, truncating beyond second-order in curved space-time in electromagnetic fields. Unlike holographic calculation~\cite{Baier:2007ix}, we show that the viscous evolution equations do not explicitly depend on the curvature of space-time. Also, we have tested the causality and stability of the second-order theory in curved space-time in the presence of linearised metric perturbation and derived dispersion relations for various modes. Interestingly, we found the coupling of gravitational modes with the usual magneto-sonic modes in the small wave-number limit. Also, we show additional non-hydrodynamical modes arise due to gravity for a bulk-viscous fluid.
	\end{abstract}
	
	\maketitle
	\section{Introduction} 
	General Relativistic Magnetohydrodynamics (GRMHD) is an essential frame-work to study astrophysical systems such as binary neutron star mergers, which is one of the source of the gravitational waves \cite{Ruiz:2020via,Cikintoglu:2022pvi,Ripperda:2019lsi,Chabanov:2021dee}. Usually, in GRMHD numerical studies, the ideal magnetohydrodynamics limit ( large magnetic Reynolds number) is used, and other dissipative effects (due to bulk and shear viscosity, etc.) are usually neglected \cite{Ripperda:2019lsi,Matthew D. Duez,Bruno Giacomazzo1,Ankan Sur,Milton Ruiz}. The ideal-MHD (note that we use "ideal" to denote large electrical conductivity, and zero or vanishing viscosity is termed as non-viscous fluid to avoid confusion) might be a good approximation for large celestial bodies where the length scale involved is large, and gradients are small~\cite{Most:2021rhr,Most:2021uck}. On the other hand, in heavy-ion collisions, two heavy nuclei collide and create a hot and dense Quark-Gluon-Plasma (QGP) that evolves for a few femtoseconds like a strongly coupled fluid, gradients are usually large, and dissipative effects are not negligible. In this case the positively charged nuclei move with relativistic speed and generate one of the strongest transient magnetic fields in the universe~\cite{Roy:2015coa,Roy:2017yvg,Skokov:2009qp,Tuchin:2013ie}. The QGP produced in the heavy-ion collisions behaves like a fluid with negligible specific shear viscosity~\cite{Chaudhuri:2012yt,Gale:2013da,Romatschke:2017ejr,Heinz:2013th}. Theoretical studies show that the QGP has finite electrical conductivity; hence the evolution of QGP in a strong electromagnetic background field naturally indicates that relativistic magnetohydrodynamics is the appropriate framework for the space-time evolution. Unlike astrophysical systems, the space-time metric considered in heavy-ion collisions is flat. In short, in GRMHD, we generally neglect the viscous effects, and in special relativistic MHD (as in heavy-ion collisions), we ignore the curvature of space-time. 
	
	Any relativistically consistent theory, in principle, must obey the causality condition, i.e., the speed of propagation of perturbations cannot be superluminal. For the case of relativistic viscous hydrodynamics, it is well known that relativistic extension of the Navier-Stokes leads to acausality \cite{Hiscock:1983zz} and one of the remedies were suggested by Mueller-Israel-Stewart (MIS)\cite{Stewart,israel_Stewart} through the inclusion of second-order dissipative corrections in entropy four-current. As we know, the second-order relativistic viscous hydrodynamics equations give rise to a causal propagation (in a restricted sense) of the perturbation in fluids~\cite{Pu:2009fj,Denicol:2008ha,Perna:2021osw,Aguilar:2017ios} and the same is true for a charged fluid in an external electromagnetic field \cite{Biswas:2020rps}. The second-order causal dissipative ideal~\cite{Denicol:2018rbw,Panda:2020zhr} and resistive \cite{Denicol:2019iyh,Panda:2021pvq} MHD formulation were recently derived from the relativistic Boltzmann kinetic equation using the moment method and the relaxation time approximations. In these MIS-type causal theories, the dissipative fluxes are treated as independent variables and are governed by relaxation-type equations. The two formulations in \cite{Denicol:2018rbw,Panda:2020zhr,Denicol:2019iyh,Panda:2021pvq} produce almost identical equations for the evolution of these dissipative quantities, albeit with different transport coefficients and some extra coefficients in RTA approximation. In both formulations, the space-time metric was assumed to be flat. In the present study, we lift this restriction, allow the space-time to be curved, and formulate the second-order dissipative resistive GRMHD. We show that this gives rise to causal theory when subjected to linear perturbations. Throughout the paper, we assume that the gravitational field varies very little over the fluid particles' mean free path and mean free time.
	
	The manuscript is arranged as follows: in Section \eqref{sec2:Boltzmann}, we describe the Boltzmann equation in the Riemannian space, where we use the relaxation time approximation for simplifying the collision kernel. In the next Section \eqref{sec3:defin}, we define various macroscopic fluid variables from the kinetic theory using a one-particle phase-space distribution function; we also discuss the conservation of energy-momentum and the Maxwell equations in the covariant form. The following Section \eqref{sec4:OrderHydro} contains the first and second-order general relativistic magnetohydrodynamics equations derived from the relativistic kinetic theory. In Section \eqref{sec:causality} we investigate the causality and stability using linear perturbation considering dynamic space-time metric. Section \eqref{sec4:OrderHydro} and 
	\eqref{sec:causality} contain the main results. Finally, we summarise our work in Section \eqref{sec:summary}.
	
	Throughout the work we use metric with mostly negative sign i.e., $g_{\mu\nu}=diag \left(+,-,-,-\right)$. We also use the natural units where
	$\hbar=c=k_{B}=\mu_0=1$. The gravitational constant is denoted by G, and the cosmological constant is denoted by $\Lambda$.
	
	\section{The Boltzmann equation in a gravitational and electromagnetic field}
	\label{sec2:Boltzmann}
	Here we give a short textbook introduction to the Boltzmann equation in Riemannian space in the presence of an external force due to  electromagnetic field. We further assume the collision integral in the Boltzmann equation is given by the relaxation time
	approximation (RTA). Though it is one of the simplest forms of the collision kernel in use, it essentially captures almost all the system's important features initially away from equilibrium and relaxing back to the nearest equilibrium state through the inter-particle collisions. For simplicity, we also assume the relaxation time is a Lorentz scalar and independent of the momentum of the particles. However, this momentum independence of the relaxation time is not a necessary condition for the RTA to be applicable \cite{Rocha:2021zcw,Mitra:2020gdk}.
	
	Now, we recall that in the Minkowski space the phase space volume element in two different inertial reference frames are related as  
	$d^3xd^3p=d^3x^{\prime}d^3p^{\prime}$. The one-particle distribution function $f(x^{\mu},p^{\mu})$ is usually written as 
	$f(\bf{x},\bf{p},t)$ due to the on shell condition $p^{0}=\sqrt{|{\bf{p}}|^2+m^2}$. In a Riemannian space the following relations hold:
	$\sqrt{-g^{\prime}}\frac{d^3p^{\prime}}{p_{0}^{\prime}}=\sqrt{-g}\frac{d^3p}{p_{0}}$, $\sqrt{-g^{\prime}}d^4x^{\prime}=\sqrt{-g}d^4x$,
	The four velocity is defined as $u^{\mu}=\Gamma(1,\bf{v})$, where the Lorentz gamma $\Gamma=\frac{1}{\sqrt{g_{00}\left(1+\frac{g_{0i}}{g_{00}}\beta^{i}\right)^2-\beta^2}}$. The covariant components are given by $u_{\mu}=g_{\mu\nu}u^{\nu}$ with the normalisation $u_{\mu}u^{\mu}=1$.
	
	Considering the Boltzmann equation in a curved space-time under an electromagnetic force, the evolution of one-particle distribution function 
	$f({\bf x}, {\bf p},t)$ is given by \cite{book_carlo}
	\begin{eqnarray}\nonumber
		&&p^{\mu}\partial_{\mu}f+qF^{\mu\nu}p_{\nu}\frac{\partial f}{\partial p^{\mu}}-\Gamma^{\mu}_{\alpha\rho}p^{\alpha}p^{\rho}\frac{\partial f}{\partial p^{\mu}}=-\frac{u\cdot p}{\tau_c}\delta f, \\\label{eq:boltzmann}
		&&p^{\mu}\partial_{\mu}f+F^{\mu}\frac{\partial f}{\partial p^{\mu}}=-\frac{u\cdot p}{\tau_c}\delta f 
	\end{eqnarray}
	where we define $F^{\mu} \equiv qF^{\mu\nu}p_{\nu}-\Gamma^{\mu}_{\alpha\rho}p^{\alpha}p^{\rho}$
	which includes the effect of gravity (which is approximated here as space-time curvature)
	and  $F^{\mu\nu}=g^{\mu\alpha}F_{\alpha\beta}g^{\beta\nu}$ is the antisymmetric electromagnetic field tensor.
	The RHS of the Boltzmann equation  Eq.\eqref{eq:boltzmann} is the collision kernel, which we have approximated to be 
	of the form of relaxation time approximation, with $\tau_c$ being the relaxation time.\\
	For a system slightly away from the local equilibrium, the one-particle distribution function 
	$f$ can be decomposed into an equilibrium  $f_0$ and an off-equilibrium part $\delta f$ as $f=f_0+\delta f$. 
	The equilibrium distribution function for particles is given by:$f_0=\frac{1}{e^{\beta\left(g_{\mu\nu } u^{\mu}p^{\nu} \right)-\alpha}+r}$,
	$r=0,\pm 1$ correspond  the boltzmann, the fermi-dirac, and the bosonic distribution respectively. For anti-particles
	$\alpha \rightarrow -\alpha$ in the above expression.\\
	At this point, we would like to take a detour and discuss about the form of the equilibrium distribution function $f_{0}$ in presence of 
	gravity. We know that for the equilibrium distribution function the RHS of the Boltzmann equation vanish by definition. The LHS of the Boltzmann equation should also vanish; this imposes some restriction on the hydrodynamic variables. If we put $f_0$ in the LHS of the Boltzmann equation we get
	\begin{equation}
		\label{eq:preKilling}
		-p^{\alpha}p^{\beta}D_{\alpha}\left(\beta u_{\beta}\right)+p^{\mu}D_{\mu}\alpha=0,
	\end{equation}
	here $D_{\alpha} u_{\beta} = \partial_{\alpha}u_{\beta} - \Gamma_{\alpha\beta}^{\mu}u_{\mu}$ is the covariant derivative of $u_\beta$.
	In Eq.\eqref{eq:preKilling} all the terms are written in terms of a polynomial equation of $p^{\mu}$, hence all the components should vanish independently. It follows that 
	\begin{eqnarray}
		D_{\mu}\alpha&=&0,
	\end{eqnarray}
	
	\[
	-D_{\alpha}(\beta u_{\beta})= 
	\begin{cases}
		0,& \text{if } m\neq 0\\
		f(x^{\delta})g_{\alpha\beta}, &\text{if }  m=0
	\end{cases}
	\]
	where $f(x^{\delta})$ is some arbitrary function of the space-time coordinate. If we consider a massless gas then $g_{\alpha\beta} p^{\alpha}p^{\beta}=0$, and for $f_0$ the LHS of the Boltzmann equation becomes zero. But for the case of $m \neq 0$ from the above equation we get :
	
	\begin{eqnarray}\nonumber
		g_{\sigma\beta}\partial_{\alpha}(\beta u^{\sigma})+g_{\sigma\alpha}\partial_{\beta}(\beta u^{\sigma})+\beta u^{\sigma}\partial_{\sigma}g_{\alpha\beta}&=&0.
	\end{eqnarray}
	
	The above equation is a Killing equation with $\beta u^{\mu}$ being the Killing vector. If we define the components of the killing vector in 
	a particular reference frame as $\beta u^{\mu}=(\beta u^0,\Vec{0})$ where $\beta u^0$ is constant, the killing equation  leads  to
	\begin{eqnarray}
		\partial_{0}g_{\alpha\beta}&=&0.
	\end{eqnarray}
	This suggests that the metric tensor must be time-independent for gas of non-vanishing rest mass to define an equilibrium distribution function. 
	However, for the massless case, there are no such restrictions. 
	
	Another point to ponder is the Liouville's theorem, which along with the on-shell condition gives rise to the following equation:
	\begin{eqnarray}
		\label{eq:Liouv}
		\frac{\partial}{\partial x^a}\left(\frac{g}{p_0}p^a \right)&=&0,
	\end{eqnarray}
	where
	\begin{eqnarray}\nonumber
		&&p^a\equiv (p^{\mu},-\Gamma^i_{\mu\nu}p^{\mu}p^{\nu})   , \\\nonumber
		&&\frac{\partial}{\partial x^a} \equiv \left(\frac{\partial}{\partial x^{\mu}}, \frac{\partial}{\partial p^i}\right).
	\end{eqnarray}
	Eq.\eqref{eq:Liouv} is useful in proving the conservation of energy-momentum and particle four-current.

	
	\section{Definitions and conservation equations}
	\label{sec3:defin}
	Now that we have defined the equilibrium distribution function and the Boltzmann equation in curved spacetime, 
	we use these definitions to define hydrodynamic variables and conservation laws that we will use later to derive the equation of motion of a fluid. The particle four flow and the energy-momentum tensors are defined as the first and second moment of the distribution function:
	\begin{eqnarray}\label{eq:defnumber}
		N^{\mu}_f &=&\int{}{}\frac{\mathcal{G}_s d^3p}{(2\pi)^3p_0}\sqrt{-g}p^{\mu}f,\\ \label{eq:defenergymomentum} 
		T^{\mu\nu}_f&=&\int{}{}\frac{\mathcal{G}_s d^3p}{(2\pi)^3p_0}\sqrt{-g}p^{\mu}p^{\nu}f.
	\end{eqnarray}
	$\mathcal{G}_s$ here is the spin-degeneracy factor.
	Usually, the particle four-current $N^{\mu}_f$ corresponds to some conserved charges, and in that case, the conservation of 
	particle and the energy-momentum tensor are obtained by replacing the partial derivatives with the covariant derivatives wherever applicable:
	\begin{eqnarray}\label{eq:connumber}
		D_{\mu}N^{\mu}_f&=&\partial_{\mu}N^{\mu}_f+\Gamma^{\mu}_{\mu\rho}N^{\rho}_f=0 ,\\\label{eq:conenergymomentum}
		D_{\mu}T^{\mu\nu}_f&=&\partial_{\mu}T^{\mu\nu}_f +\Gamma^{\mu}_{\mu\rho}T^{\nu\rho}_f +\Gamma^{\nu}_{\mu\rho}T^{\mu\rho}_f =0,
	\end{eqnarray}
	here $\Gamma^{\mu}_{\nu\rho}$ is the Christoffel symbol of second kind and is given by :
	$\Gamma^{\mu}_{\nu\rho}=\frac{g^{\mu\lambda}}{2}\left(\partial_{\nu}g_{\rho\lambda}+\partial_{\rho}g_{\nu\lambda}-\partial_{\lambda}g_{\nu\rho}\right)$ , $u\cdot p$=$g_{\mu\nu}u^{\mu}p^{\nu}$ , and $g_{\alpha\beta}u^{\alpha}u^{\beta}=1$.\\
	Also, we use $D=u^{\mu}D_{\mu}$ for the covariant time derivative, and  the covariant derivative is decomposed as  $D_{\mu}=u_{\mu}D+\nabla_{\mu}$, where $\nabla_{\mu}=\Delta^{\alpha}_{\mu}D_{\alpha}$. We decompose the gradient of four velocity in terms of 
	symmetric traceless shear, non-zero trace bulk, and antisymmetric second rank vorticity tensor as 
	$\nabla_{\mu}u_{\nu}=\sigma_{\mu\nu}+\Delta_{\mu\nu}\frac{\theta}{3}+\omega_{\mu\nu}$  where 
	$\Delta_{\mu\nu}=g_{\mu\nu}-u_{\mu}u_{\nu}$ ,$\theta=D_{\mu}u^{\mu}$, $D_{\mu}g^{\mu\nu}=0$.
	The electromagnetic field tensor is given by 
	$F_{\mu\nu}=D_{\mu}A_{\nu}-D_{\nu}A_{\mu}=E_{\mu}u_{\nu}-E_{\nu}u_{\mu}+B_{\mu\nu}$,
	where $B_{\mu\nu}=\epsilon_{\mu\nu\alpha\beta}u^{\alpha}B^{\beta}$,
	$B^{\mu\nu}=-B b^{\mu\nu}$, $\epsilon_{\mu\nu\alpha\beta}=\sqrt{-g}$ , and $\epsilon^{\mu\nu\alpha\beta}=\frac{-1}{\sqrt{-g}}$.
	Since we consider a combined system of a fluid , electromagnetic fields, and the presence of a conserved charged current, the corresponding 
	energy-momentum tensor of the fluid, the EM fields, and the four current of the conserved quantity are given by:
	
	\begin{eqnarray}\label{eq:energy-momentum_fluid}
		T^{\mu\nu}_f&=&\epsilon u^{\mu}u^{\nu}-\left(P+\Pi \right)\Delta^{\mu\nu}+\pi^{\mu\nu}, \\ \label{eq:energy-momentum_field}
		T^{\mu\nu }_{EM}&=&-F^{\mu}_{\alpha}F^{\nu\alpha}+\frac{1}{4}g^{\mu\nu}F_{\alpha\beta}F^{\alpha\beta},\\ 
		N^{\mu}_f&=&n_f u^{\mu}+V_f^{\mu}.
	\end{eqnarray}
	
	The fluid though under consideration is non-polarisable and non-magnetisable.
	While deriving the conservation equations from the  Boltzmann equation Eq.\eqref{eq:boltzmann} we want to express the left hand side 
	in terms of thermodynamic forces (spatial gradients of velocities, temperature etc.). For that we note $\partial_{\lambda}f_{0}=-f_{0}\tilde{f}_{0}\left[\beta p^{\mu}D_{\lambda}u_{\mu}+\beta u_{\mu}\Gamma^{\mu}_{\lambda\rho}P^{\rho} -\partial_{\lambda}\alpha+(p\cdot u)\partial_{\lambda}\beta \right]$ (where we have used $\tilde{f}_{0}=(1-r f_{0})$, $r=\pm1$ and $ \partial_{\lambda}p^{\nu}=0$), which gives rise to terms such as 
	$\dot{\alpha}$, $\dot{\beta}$, $\dot{u_{\mu}}$. Such terms do not contribute in the expression for the thermodynamic force, hence we need to eliminate them. This we achieve with the help of the conservation equations given below along with the relativistic thermodynamic integrals  $J^{(m)\pm}_{nq}$ defined later in appendix\eqref{app:definition}.
	\begin{eqnarray}\nonumber
		D_{\mu}T^{\mu\nu}_{EM}&=&-F^{\nu\lambda}J_{\lambda f},\\ \nonumber
		D_{\mu}J^{\mu}&=&0,\\\nonumber
		T^{\mu\nu}&=&T^{\mu\nu}_f + T^{\mu\nu}_{EM},\\ \nonumber
		D_{\mu}T^{\mu\nu}_f&=& F^{\nu\lambda}J_{\lambda f}.
	\end{eqnarray}
	Now using the fluid conservation equation, the expressions for $\theta$ , $D$, $\sigma_{\mu\nu}$,  and $\omega_{\mu\nu}$ 
	we get the following expressions for $\dot{\alpha}$, $\dot{\beta}$ and $\dot{u}^{\mu}$ :\\
	\begin{widetext}
		\begin{eqnarray}
			\dot{\alpha}&=&\frac{1}{D_{20}}\left[J_{20}^{(0)-}\theta\left(\epsilon +P +\Pi\right)-J_{30}^{(0)+}\left(n_f \theta +D_{\mu}V_f^{\mu} \right) +J_{20}^{(0)-}\left(-\pi^{\mu\nu}\sigma_{\mu\nu}+qE^{\mu}V_{f\mu}\right)\right],
			\nonumber
			\\
			\dot{\beta}&=&\frac{1}{D_{20}}\left[J_{10}^{(0)+}\theta\left(\epsilon +P +\Pi\right) -J_{20}^{(0)-}\left(n_f \theta +D_{\mu}V_f^{\mu} \right) +J_{10}^{(0)+}\left(-\pi^{\mu\nu}\sigma_{\mu\nu}+qE^{\mu}V_{f\mu} \right) \right],
			\nonumber
			\\ \nonumber
			\dot{u}^{\mu}&=&\frac{1}{\epsilon+P}\left[\frac{n_f}{\beta}\left(\nabla^{\mu}\alpha-h \nabla^{\mu}\beta\right)-\Pi \dot{u}^{\mu}+\nabla^{\mu}\Pi-\Delta^{\mu}_{\nu}D_{\rho}\pi^{\rho \nu}\right] +\frac{1}{\epsilon+P}\left[q n_f E^{\mu}-qB b^{\mu\nu}V_{f \nu} \right],
		\end{eqnarray}
	\end{widetext}	
	
	where $D_{20}=J^{(0)+}_{30}J^{(0)+}_{10}-J_{20}^{(0)-}J_{20}^{(0)-}$, $h=\frac{\epsilon+P}{n_f}$ and $\sigma^{\mu\nu}=\Delta^{\mu\nu}_{\alpha\beta}\nabla^{\alpha}u^{\beta}$. These relations will be used later to derive the off-equilibrium distribution function $\delta f$.
	
	Now the definition of viscous stresses given in Eq.(\ref{eq:energy-momentum_fluid}) in terms of $\delta f$ are given by:
	\begin{eqnarray}
		\label{eq:pishear}
		\pi^{\mu\nu}&=&\Delta^{\mu\nu}_{\alpha\beta}\int \frac{\mathcal{G}_s d^3p \sqrt{-g}}{(2\pi)^3 p_0}p^{\alpha}p^{\beta}(\delta f),\\
		\label{eq:pibulk}
		\Pi&=&-\frac{\Delta_{\alpha\beta}}{3}\int \frac{\mathcal{G}_s d^3p \sqrt{-g}}{(2\pi)^3 p_0}p^{\alpha}p^{\beta}(\delta f),\\
		\label{eq:diff}
		V^{\mu}&=&\Delta^{\mu}_{\alpha}\int \frac{\mathcal{G}_s d^3p \sqrt{-g}}{(2\pi)^3 p_0}p^{\alpha}(\delta f).
	\end{eqnarray}

	\section{Order-by-order magnetohydrodynamic equations}
	\label{sec4:OrderHydro}	
	\subsection{Calculation of $\delta f$}
	
	In this section we discuss the details of $\delta f$ which is used in Eqs. \eqref{eq:pishear},\eqref{eq:pibulk},\eqref{eq:diff} to derive the 
	evolution equation for dissipative stresses upto second order. As mentioned earlier, we use RTA collision kernel in the Boltzmann 
	equation to derive $\delta f$ as a series expansion around $f_{0}$ in terms of three smallness paramaters Knudsen number $\mathrm{Kn} = \tau_c \partial_{\mu}$~\cite{Jaiswal:2016pmi} , $\chi = qB\tau_c /T$ ,$\xi=qE\tau_c/T$ \cite{Panda:2020zhr,Panda:2021pvq,Mohanty:2018eja}. 
	We note that this is not the unique way to derive the $\delta f$, for example in \cite{Denicol:2018rbw,Mitra:2019jld} a relativistic moment  method was used, and some other variant of RTA can be found in \cite{Ghiglieri:2018dgf,Arnold:2000dr,Arnold:2003zc}.
	In addition due to non-zero Christoffel symbols we also need to make sure that $\tau_c \Gamma^{\mu}_{\alpha\beta}<<1$ holds (corresponds to nearly flat geometry  between two collision)which corresponds to a weak gravitaional force limit, $\mathcal{F}_g^{\mu} << T^3$, where $\mathcal{F}_g^{\mu} =\Gamma^{\mu}_{\alpha\beta} p^{\alpha}p^{\beta} $. With the above assumptions we are ready to derive the first order correction $\delta f^{(1)}$ by noting :
	$f=f^{(0)}+\delta f^{(1)}+\delta f^{(2)}+...$, where $\delta f^{(n)}$ is obtained from the Boltzmann equation 
	\begin{eqnarray}\nonumber
		&&p^{\mu}\partial_{\mu}f+qF^{\mu\nu}p_{\nu}\frac{\partial f}{\partial p^{\mu}}-\Gamma^{\mu}_{\alpha\rho}p^{\alpha}p^{\rho}\frac{\partial f}{\partial p^{\mu}}=-\frac{u.p}{\tau_c}\delta f ,\\ \nonumber
	\end{eqnarray}
	which leads to 
	\begin{equation}
		\label{eq:deltafn}
		\delta f^{(n)}=\sum_{n=0}^{\infty}{\left[-\frac{\tau_c}{u. p}\left(p^{\mu}\partial_{\mu}+\mathcal{F}^{\mu}\frac{\partial}{\partial p^{\mu}}-\mathcal{F}_g^{\mu} \frac{\partial}{\partial p^{\mu}}\right)\right]}^nf_0,
	\end{equation}
	where $\mathcal{F}^{\mu}\equiv qF^{\mu\nu}p_{\nu}$.
	The detailed derivation and explicit expression for $\delta f^{(1)}$  and $\delta f^{(2)}$ are given in appendix (\ref{app:2}). 
	
	\begin{widetext}
		Using the expression for $D_{\mu}u_{\alpha}$( see Eq.(\ref{app:covU})) in the expression for $\delta f^{(1)}$ (Eq.(\ref{app:deltaf1})) we have 
		\begin{eqnarray}\nonumber
			\delta f^{(1)}&=&f_0 \tilde{f}_0\frac{\tau_c}{u.p}\left((u.p)p^{\mu}\partial_{\mu}\beta +\beta p^{\mu}p^{\alpha}\left(u_{\mu}\dot{u}_{\alpha}+\omega_{\mu\alpha}+\sigma_{\mu\alpha}+\frac{\Delta_{\mu\alpha}\theta}{3}\right)-p^{\mu}\partial_{\mu}\alpha\right)\\ \label{eq:deltaf1}
			&&+f_0 \tilde{f}_0\frac{\tau_c}{u.p}{\beta qF^{\mu\nu}\delta^{s}_{\mu}u_{s}p_{\nu}}.
		\end{eqnarray}
		Similarly we can calculate $ \delta f^{(2)}$:
		\begin{eqnarray}\label{eq:deltaf2}
			\delta f^{(2)}&=&-\frac{\tau_c}{u.p}\left(p^{\mu}\partial_{\mu}  \delta f^{(1)}+qF^{\mu\nu}p_{\nu}\frac{\partial \delta f^{(1)}}{\partial p^{\mu}}-\Gamma^{\mu}_{\nu\rho}p^{\nu}p^{\rho}\frac{\partial \delta f^{(1)}}{\partial p^{\mu}}\right).
		\end{eqnarray}
		The explicit expression is quite long and hence given in appendix(\ref{app:2}). 
	\end{widetext}

	\subsection{First Order Constitutive relations}
	Using $\delta f^{(1)}$ we get the first-order constitutive relations for dissipative stresses as,
	\begin{eqnarray} \label{eq:1bulk}
		\Pi_{(1)}&=&-\tau_c \beta_{\Pi}\theta,\\ \label{eq:diffusion}
		V^{\mu}_{(1)}&=&\tau_c \beta_V \left(\nabla^{\mu}\alpha+ \beta  q E^{\mu}\right),\\  \label{eq:shear}
		\pi^{\mu\nu}_{(1)}&=&2\tau_c \beta_{\pi}\sigma^{\mu\nu}.
	\end{eqnarray}
	
	where  $\beta_{\Pi}=\frac{5\beta}{3}J_{42}^{(1)+} +  \mathcal{X} J_{31}^{(0)+}-\mathcal{Y} J_{21}^{(0)-}$ with $$\mathcal{X}=\frac{J_{10}^{(0)+}\left( \epsilon+P\right)-J_{20}^{(0)-}n_f}{D_{20}},$$ and $$\mathcal{Y}=\frac{J_{20}^{(0)-}\left( \epsilon+P\right)-J_{30}^{(0)+}n_f}{D_{20}}.$$, $\beta_V =\frac{1}{h}J_{21}^{(0)-}-J_{21}^{(1)-}$ and $\beta_{\pi}=\beta J_{42}^{(1)+}$.
	Not surprisingly, the results are similar to that of \cite{Panda:2021pvq} apart from the fact that the partial derivatives are now replaced by covariant derivative.
	\begin{widetext}

		\subsection{Second order evolution equations}
		In a similar manner using the expression of $\delta f ^{(2)}$ we find the second-order evolution equation,
		\begin{eqnarray}
			\frac{\Pi}{\tau_c}&=&-\dot{\Pi}-\delta_{\Pi\Pi}\Pi \theta +\lambda_{\Pi\pi}\pi^{\mu\nu}
			\sigma_{\mu\nu}-\tau_{\Pi V}V\cdot \dot{u}-\lambda_{\Pi V}V\cdot \nabla \alpha -l_{\Pi V}D \cdot V-\beta_{\Pi}\theta-
			qB\lambda_{\Pi VB} b^{\mu\beta}V_{\beta}V_{\mu}
			\nonumber
			\\
			&&+\tau_c \tau_{\Pi V B}\dot{u}_{\alpha} qBb^{\alpha\beta}V_{\beta}- q
			\delta_{\Pi V B}\nabla_{\mu}\left(\tau_c B b^{\mu\beta}V_{\beta} \right)  -q^2\tau_{c}\chi_{\Pi EE}E^{\mu}E_{\mu},
			\label{eq:2bulkevolutionexp} \\ 
			\frac{V^{\mu}}{\tau_c}&=&-\dot{V}^{\langle\mu\rangle}-V_{\nu}\omega^{\nu \mu}
			+\lambda_{VV}V^{\nu}\sigma^{\mu}_{\nu}-\delta_{VV}V^{\mu} \theta
			+\lambda_{V\Pi}\Pi\nabla^{\mu} \alpha -\lambda_{V\pi}\pi^{\mu \nu}\nabla_{\nu}\alpha
			-\tau_{V\pi}\pi^{\mu}_{\nu}\dot{u^{\nu}}-q B \delta_{V B} b^{\mu\gamma}V_{\gamma} \nonumber
			\\
			&& +\tau_{V\Pi}\Pi \dot{u^{\mu}}+l_{V\pi}\Delta^{\mu \nu} D_{\gamma}\pi^{\gamma}_{\nu}
			-l_{V\Pi}\nabla^{\mu}\Pi+\beta_V \nabla^{\mu} \alpha +\tau_c q B l_{V\pi B} b^{\sigma \mu} D^{\kappa}\pi_{\kappa\sigma}-q \tau_c \lambda_{VVB}B
			b^{\gamma \nu}V_{\nu}\sigma^{\mu}_{\gamma} \nonumber \\
			&& +\tau_c q B \tau_{V \Pi B}b^{\gamma\mu}\Pi \dot{u}_{\gamma} -\tau_c qB l_{V \Pi B} b^{\gamma\mu}
			\nabla_{\gamma}\Pi-q\tau_c \delta_{VVB} B b^{\mu\nu}V_{\nu}\theta -q \tau_c \mathbf{\rho}_{VVB} B b^{\gamma \nu}V_{\nu}
			\omega^{\mu}_{\gamma} 
			\nonumber \\
			&&+ \chi_{VE}  q E^{\mu}+q\Delta^{\mu}_{\alpha}\chi_{VE}D\left(\tau_c E^{\alpha}\right)-q \tau_c \rho_{VE} E^{\mu}\theta	- q \tau_{VVB}\Delta^{\mu}_{\gamma}D\left(\tau_cBb^{\gamma \nu}V_{\nu} \right),
			\label{eq:fulldiffusion} \\ 
			\frac{\pi^{\mu\nu}}{\tau_c}&=&-\dot{\pi}^{\left<\mu\nu\right>}+2\beta_{\pi}\sigma^{\mu\nu}
			+2\pi^{\langle\mu}_{\gamma}\omega^{\nu\rangle\gamma}-\tau_{\pi\pi}
			\pi^{\langle\mu}_{\gamma}\sigma^{\nu\rangle\gamma} -\delta_{\pi\pi}\pi^{\mu\nu}\theta  +\lambda_{\pi\Pi}\Pi\sigma^{\mu\nu}-\tau_{\pi V}V^{\langle\mu}\dot{u}^{\nu\rangle}-\tau_c qB\tau_{\pi VB} \dot{u}^{\langle\mu}b^{\nu\rangle\sigma} V_{\sigma}
			\nonumber \\
			&&+\lambda_{\pi V}V^{\langle\mu}\nabla^{\nu \rangle}\alpha -l_{\pi V}\nabla^{\langle\mu}
			V^{\nu\rangle} +\delta_{\pi B}\Delta^{\mu\nu}_{\eta \beta}q B b^{\gamma \eta}g^{\beta \rho}
			\pi_{\gamma\rho} - qB\lambda_{\pi VB} V_{\gamma}b^{\gamma\langle\mu}V^{\nu\rangle}
			-q \delta_{\pi VB}   \nabla^{\langle\mu}\left(\tau_c B^{\nu\rangle\gamma}V_{\gamma} \right) \nonumber
			\\
			&&+ q^2\tau_{c} \chi_{\pi EE} \Delta^{\mu\nu}_{\sigma \rho } E^{\sigma}E^{\rho}. 
			\label{eq:2evolutionexp}
		\end{eqnarray}
		Here the evolution equations are similar to that found in~\cite{Panda:2021pvq} apart from the fact that the partial derivatives are now replaced by covariant derivatives. This shows no explicit dependence on the space-time curvature (if present) as was also inferred in~\cite{Baier:2007ix} and as was taken in~\cite{Most:2021uck} . However, the explicit expression given in the appendix \eqref{app:2} might be helpful for numerical simulations for a given geometry.
	\end{widetext}

	\section{Causality and stability for GRMHD}
	\label{sec:causality}
	
	Here we study the causality and stability of non-resistive-MHD formulation in presence of curved space-time and small metric perturbation. For convenience we decompose the perturbation in the fluid variables and the space-time metric in Fourier modes of the form $ {\delta \tilde{X}}= \delta X e^{ig_{\mu\nu}k^{\mu}x^{\nu}}$, where $\delta X \equiv (\delta \epsilon, \delta P,\delta u^{\mu}..)$. We decompose the metric as  $g_{\mu\nu}= \eta_{\mu\nu} + \delta \tilde{g}_{\mu\nu}$, where $\eta_{\mu\nu}$ is the flat space-time Minkowski metric, and $\delta \tilde{g}_{\mu\nu}$  is the small perturbation. Due to the fact that $g^{\mu\alpha}g_{\alpha\nu}=\delta^{\mu}_{\nu}$ we have 
	\begin{eqnarray}\nonumber
		\delta [g^{\mu\alpha} g_{\alpha\nu}] &=& 0 \\ \nonumber
		\delta \tilde{g}^{\mu\nu}&=& \delta \tilde{g}_{\alpha\beta} \eta^{\alpha\mu} \eta^{\beta\nu}.
	\end{eqnarray}
	
	Before we linearise the conservation equations for fluid, let us discuss the linearised form of the Einstein equation that is given by :
	
	\begin{eqnarray}\nonumber
		\delta \tilde{R}_{\mu\nu} -\frac{1}{2}  \eta_{\mu\nu } \eta^{\alpha\beta} \delta \tilde{R}_{\alpha\beta}  + \Lambda \delta \tilde{g}_{\mu\nu} &=&- 8\pi G \delta \tilde{T}_{\mu\nu}, \\ 
		\label{eq:linEinst1}
		-\partial_{\nu} \delta \tilde{\Gamma}^{\lambda}_{\mu\lambda}+\partial_{\lambda} \delta \tilde{\Gamma}^{\lambda}_{\mu\nu} + \Lambda  \delta \tilde{g}_{\mu\nu} &=& -8\pi G \delta \tilde{T}_{\mu\nu}.
	\end{eqnarray}
	We use the explicit form of $\delta \tilde{\Gamma}^{\mu}_{\nu\alpha}$ (the details are given appendix\eqref{app:deltagamma}) in Eq.\eqref{eq:linEinst1} to  derive Eq.\eqref{eq:deltagXX}, Eq.\eqref{eq:deltagXY}.

	
	For the sake of simplicity, without losing generalisation, one can use Lorentz gauge $\delta \tilde{g}^{\mu\nu} k_{\nu} =0$ along with transverse $\delta \tilde{g}^{\mu\nu} u_{\nu}=0$, and traceless gauge $\delta \tilde{g}^{\mu}_{\mu}=0$ to reduce the number of independent variables. With the following choice for $k^{\mu} = \left(\omega , k sin \theta , 0 , k cos \theta\right)$ , and $u^{\mu}= (1,0,0,0)$ we have the following form for the above mentioned three gauges:
	\begin{eqnarray}
		\delta \tilde{g}^{\mu t} &=& 0, \\ 
		\delta \tilde{g}^{\mu x} k_x + \delta \tilde{g}^{\mu z} k_z&=& 0 , \\ 
		\ \delta \tilde{g}^{tt} -  \delta \tilde{g}^{xx} -  \delta \tilde{g}^{yy} -  \delta \tilde{g}^{zz} &=&0. 
	\end{eqnarray}
	
	For this case we have $\delta \tilde{g}^{xx}$ and $\delta \tilde{g}^{xy}$ as two independent components;
	hence from Eq.\eqref{eq:linEinst1}  we get the following equations for $\delta \tilde{g}^{xx}$ and  $\delta \tilde{g}^{xy}$:
	\begin{eqnarray}\nonumber
		&&- \left(\Lambda  + \frac{1}{2}   \left( \omega^2 - k^2 \right) - 8 \pi G \left( P_0  + \frac{B^2_0}{2} \right)\right) \delta \tilde{g}^{xx} = \\ \label{eq:deltagXX}
		&& - 8\pi G \left( c_s^2 \delta \tilde{\epsilon} + \delta \tilde{\Pi} +   \delta \tilde{b}^z  B^2_0  - \delta \tilde{g}^{xx} \frac{(k_x)^2}{(k_z)^2} \frac{B^2_0}{2}  \right)   \\  
		\label{eq:deltagXY}
		&&\left( \Lambda +\frac{\left( \omega^2 - k^2 \right) }{2}  
		-8 \pi G \left(P_0  + \frac{B^2_0}{2} \right)\right) \delta \tilde{g}^{xy}=0. 
	\end{eqnarray}
	
	Here we consider only bulk viscosity in the $\delta \tilde{T}^{\mu\nu}$. Next, we linearise the Maxwell's equation $D_{\mu} \tilde{F}^{\mu\nu} = 0$  by taking $b^{\mu}\equiv B^{\mu}/B = (0,0,0,1)$ . Also note that we have considered $\Pi_0 =0$.

	The resulting  linearised equations are  :
	\begin{eqnarray}
		i k_{x} \delta \tilde{b}^{x}     + i k_{z}  \delta \tilde{b}^{z}         &=& 0,  \\
		ik_z   \delta \tilde{u}^{x}    -  i\omega   \delta \tilde{b}^{x}    &=& 0,  \\ 
		i k_z   \delta \tilde{u}^{y}   - i\omega  \delta \tilde{b}^{y}   &=& 0,  \\ 
		-i \omega    \delta \tilde{b}^{z}   -  ik_{x} \delta \tilde{u}^{x}   &=& 0.   
	\end{eqnarray}

	Next, we linearise the energy-momentum conservation equation for the fluid and field:
	\begin{eqnarray}
		\partial_{\mu} \delta \tilde{ T}^{\mu\nu} + \delta \tilde{\Gamma}^{\mu}_{\mu\alpha} T^{\alpha\nu}_0 + \delta \tilde{\Gamma}^{\nu}_{\mu\alpha} T^{\mu\alpha}_0 &=& 0 \\ \nonumber
	\end{eqnarray}
	\begin{widetext}
		The resulting linearised equations are  :
		\begin{eqnarray}\label{fluid1}
			&& i \omega \delta \tilde{\epsilon} - i \omega\delta \tilde{\Pi}  +  i k_x  \left(\epsilon_0  + P_0  \right)    \delta \tilde{u}^{x}  + ik_z \left(\epsilon_0  + P_0   \right)  \delta \tilde{u}^{z}
			=0, \\ \label{fluid2}
			&&i \omega   \left(\epsilon_0 + P_0 + B^2_0-    \frac{k_x^2}{\omega^2} B^2_0 \right)  \delta \tilde{u}^{x}    
			+ i k_x  c_s^2 \delta \tilde{\epsilon}     + i k_x \delta \tilde{\Pi}  
			- ik_x \frac{k_x^2 + k_z^2}{k_z^2} \delta   B_0^2      \tilde{g}^{xx}       -  ik_z B^2_0  \delta \tilde{b}^{x}
			=0, \\ \label{fluid3}
			&&i \omega   \left(\epsilon_0 + P_0 + B^2_0 \right) \delta \tilde{u}^{y}    
			- ik_{z}  B^2_0 \delta \tilde{b}^{y} 
			-   ik_x  B^2_0  \delta \tilde{g}^{xy}       
			=0,  \\ \label{fluid4}
			&& i \omega   \left(\epsilon_0 + P_0  \right) \delta \tilde{u}^{z}     + i k_z   c_s^2 \delta \tilde{\epsilon}    +i k_z \delta \tilde{\Pi}   +      \frac{3ik_z k_x}{\omega}  B^2_0 \delta \tilde{u}^x   + \frac{ik_x^2 }{k_z} B^2_0  \delta \tilde{g}^{xx}  - ik_x B^2_0 \delta \tilde{b}^{x}   =0.
		\end{eqnarray} 
	\end{widetext}

	We note that bulk stress in the present formulation is an independent dynamical variable; linearising the bulk evolution equation we get  :
	\begin{eqnarray}
		\delta \tilde{\Pi} \left(  \frac{1}{\tau_c} + i \omega \right) + \beta_{\Pi} ik_{x} \delta \tilde{u}^{x}  + \beta_{\Pi} ik_{z} \delta \tilde{u}^{z}  = 0.
	\end{eqnarray}
	
	Now we have all the ingredients to derive the dispersion relations and study the causality and stability of the system under consideration. 
	For clarity, we consider three different cases one by one, as discussed below.
	
	\subsection{Non-viscous fluid with Gravity} \label{Ideal fluid with Gravity:}
	The first case we study is for non-viscous fluid with metric perturbation. Here the independent variables are  $X\equiv \left( \delta \tilde{\epsilon} , \delta \tilde{u}^x ,\delta \tilde{u}^y , \delta \tilde{u}^z , \delta \tilde{g}^{xx} , \delta \tilde{g}^{xy}   \right)^T$, the superscript $T$ denotes the transpose.  Using Eqs.(\eqref{eq:deltagXX},\eqref{eq:deltagXY}) and Eqs.(\ref{fluid1},\ref{fluid2},\ref{fluid3},\ref{fluid4}) we obtain $\mathcal{A}X=0$ where the matrix $\mathcal{A}$ is given below.  We set $\Pi=0$, and $B_0=0$ while constructing $\mathcal{A}$.	
	
	\begin{eqnarray}
		\mathcal{A}=\left[ {\begin{array}{ccccccc}
				\begin{smallmatrix}
					i \omega & (\epsilon_0 + P_0 ) ik_x & 0 & (\epsilon_0 +  P_0) ik_z & 0 & 0  \\ \nonumber 
					i k_x c_s^2 & (\epsilon_0 +  P_0 )  i \omega & 0 & 0 & 0 & 0  \\ \nonumber
					0 & 0 & (\epsilon_0 +  P_0 )  i \omega & 0 & 0 & 0  \\ \nonumber
					ik_z c_s^2 & 0 & 0 &  (\epsilon_0 + P_0 )  i \omega & 0 & 0  \\ \nonumber
					8 \pi G c_s^2 & 0 & 0 & 0 & \mathcal{M} & 0  \\ \nonumber
					0 & 0 & 0 & 0 & 0 & \mathcal{M} \\ \nonumber 
				\end{smallmatrix}
		\end{array} } \right].
	\end{eqnarray}
	
	Here we defined  $\mathcal{M} = - \left(\Lambda + \frac{1}{2} (\omega^2 - k^2) - 8 \pi G P_0 \right)$ for brevity.
	Setting det($\mathcal{A}$)=0 we obtain the dispersion relations. A careful analysis shows that in this case we have two different modes of
	wave propagation namely (i) sound modes with the dispersion relation $\omega = \pm c_s \sqrt{k_x^2 + k_z^2}$, and (ii) the gravitational modes with the dispersion relation $\omega = \pm \sqrt{k^2 - 2 \Lambda +16  G \pi P_0 }$ , where $k = \sqrt{k_x^2 + k_z^2}$.
	As expected the group velocity for the sound mode is $v_g = \frac{\partial \omega}{\partial k} = \pm c_s$, whereas for the gravitation mode $v_g = \frac{\partial \omega}{\partial k} = \pm \frac{  k }{ \sqrt{k^2 - 2 \Lambda + 16  G \pi P_0 }} $ . Here we observe a few points
	from the group velocity of the gravitational mode: since the term inside the square root in the denominator may become negative for 
	$k< \pm \sqrt{(2\Lambda - 16\pi G P_0 )}$;  we do not have any propagating modes for $k$ smaller than this cutoff and one of the roots will give rise to a unstable mode. We also note that for $k$ just above this cutoff value may results in $v_g > 1.$ However, note that we should actually consider $\lim_{k\rightarrow \infty}\frac{d\omega}{dk} = 1$ for checking the causality \cite{Pu:2009fj}; which for this case shows that the  gravitational wave travels with speed of light in vacuum. These are expected results as it is known that in non-viscous fluid the disturbance propagates with the speed of sound $c_s$ and the gravitational waves propagates with the speed of light. Now let us turn to the second case where we include bulk viscosity in the fluid.


	\subsection{ Bulk viscosity with Gravity}{\label{ Bulk viscosity with Gravity:}}
	
	When we add bulk viscosity to the fluid we have the following set of independent variables $X\equiv \left( \delta \tilde{\epsilon} , \delta \tilde{u}^x ,\delta \tilde{u}^y , \delta \tilde{u}^z , \delta \tilde{\Pi}, \delta \tilde{g}^{xx} , \delta \tilde{g}^{xy}   \right)^T$. In this case, the matrix $\mathcal{A}$
	takes the following form:

	\begin{equation}
		\mathcal{A}=\left[ {\begin{array}{cccccccc}
				\begin{smallmatrix}
					i\omega & ( \epsilon_0 + P_0) ik_x & 0 & ( \epsilon_0 + P_0 )  ik_z & -i\omega & 0 & 0  \\ \nonumber
					ik_x c_s^2 & (\epsilon_0 + P_0)  i\omega & 0 & 0 & ik_x & 0 & 0  \\ \nonumber
					0 & 0 & (\epsilon_0 + P_0)  i\omega & 0 & 0 & 0 & 0  \\ \nonumber
					ik_z c_s^2 & 0 & 0 & (\epsilon_0 + P_0)  i\omega & ik_z & 0 & 0  \\ \nonumber
					0 & ik_x \beta_{\Pi} & 0 & ik_z \beta_{\Pi} & (\frac{1}{\tau_c} + i\omega) & 0 & 0  \\ \nonumber
					8 \pi G c_s^2 & 0 & 0 & 0 &  8 \pi G & \mathcal{M} & 0 \\ \nonumber
					0 & 0 & 0 & 0 & 0 & 0 & \mathcal{M}  \\ \nonumber
				\end{smallmatrix}
		\end{array} } \right].
	\end{equation}

	Here $\mathcal{M} = - \left(\Lambda + \frac{1}{2} (\omega^2 - k^2) - 8 \pi G P_0 \right)$. Like the previous case here also
	we notice that the gravitational modes are decoupled from the hydrodynamic modes. The dispersion relation for the gravitational waves 
	are $\omega = \pm \sqrt{k^2 - 2 \Lambda +16  G \pi P_0 } $. The full dispersion relations for the hydro modes are quite long and
	hence we show here only the small and large $k$ limits.\\
	For small $k$ limit we have : \\
	$ \omega = \begin{cases} \frac{i}{\tau_c} + \sqrt{\frac{\beta_{\Pi}}{2 \epsilon_0}} k + \mathcal{O}(k^2), \\
		\pm c_s k + \mathcal{O} (k^2). \end{cases}$
	
	Here the first expression corresponds to a non-hydro mode and the second expression corresponds to the usual sound mode with speed
	of sound $c_s$.		
	On the other hand, for the large k limit we have $\omega = \pm v_g k$ where $v_g = \pm \sqrt{\frac{\beta_{\Pi} + P_0}{\epsilon_0}}$. Thus for causality to hold we must have $ \beta_{\Pi}\leq (1-c_s^2)\epsilon_0$. Since for this case $\Im{(\omega)} > 0$ all modes are stable. For the gravitational sector, we have exactly the same conclusion as was derived in the previous sub-section\eqref{Ideal fluid with Gravity:}.

	\subsection{ Bulk viscosity , magnetic field, and Gravity }\label{subsec:BulkMagneticGravity}
	Finally, in this section, we consider a fluid with finite bulk viscosity in a magnetic field and we allow small metric perturbation.
	A straightforward extension of the previous cases give the following characteristic matrix:
	
	\begin{widetext}
		\[
		\mathcal{A}=\left[ {\begin{array}{cccccccccc}
				\begin{smallmatrix}
					i\omega & (\epsilon_0 + P_0 )  ik_x & 0 & (\epsilon_0 + P_0)  ik_z & -i\omega & 0 & 0 & 0 & 0  \\ \nonumber
					i k_x c_s^2 & (\epsilon_0 + P_0 + B_0^2)i \omega -i \frac{k_x^2}{\omega} B_0^2 & 0 & 0 & ik_x & -ik_z B_0^2 & 0 & -ik_x B_0^2 \frac{k_x^2 + k_z^2}{k_z^2} & 0  \\ \nonumber
					0 & 0 & i\omega (\epsilon_0 + P_0 + B_0^2) & 0 & 0 & 0 & -ik_z B^2_0 & 0 & -ik_x B^2_0 \\ \nonumber
					ik_z c_s^2 & i\frac{3 k_x k_z}{\omega} B_0^2 & 0 & i\omega (\epsilon_0 + P_0 ) & ik_z & -ik_x B^2_0 & 0 & -ik_z B^2_0 + i k_z B_0^2 \frac{k^2}{k_z^2} & 0  \\ \nonumber
					0 & ik_x \beta_{\Pi} & 0 & ik_z \beta_{\Pi} & (\frac{1}{\tau_c} + i\omega) & 0 & 0 & 0 & 0 \\ \nonumber
					0 & ik_z  & 0 & 0 & 0 & -i\omega  & 0 & 0 & 0  \\ \nonumber
					0 & 0 & ik_z  & 0 & 0 & 0 & -i\omega  & 0 & 0  \\ \nonumber
					8 \pi G c_s^2  & \frac{-8 \pi G k_x}{\omega} B_0^2 & 0 & 0 & 8\pi G & 0 & 0 & \mathcal{N} & 0  \\ \nonumber
					0 & 0 & 0 & 0 & 0 & 0 & 0 & 0 & \mathcal{S}  \\ \nonumber 
				\end{smallmatrix}
		\end{array} } \right],
		\]
		
	\end{widetext}		
	where $\mathcal{N}= - \left( \Lambda + \frac{\omega^2- k^2}{2} - 8\pi G \left( P_0 + B_0^2- \frac{B_0^2}{2} \frac{k^2 }{k_z^2} \right) \right)$, $\mathcal{S}= - \left( \Lambda + \frac{\omega^2- k^2}{2} - 8\pi G P_0 - 8\pi G B_0^2/2\right)$, for this case the vector $X$ is $\left( \delta \tilde{\epsilon} , \delta \tilde{u}^x ,\delta \tilde{u}^y , \delta \tilde{u}^z , \delta \tilde{\Pi}, \delta \tilde{b}^x , \delta \tilde{b}^y, \delta \tilde{g}^{xx} , \delta \tilde{g}^{xy}   \right)^T$. We checked the dispersion relation and found that the usual Alfven modes
	due to finite magnetic field has the following form $\omega = \pm \frac{B_0 k_z}{ \sqrt{B_0^2 + \epsilon_0 + P_0 }}$ , where $k_z = kcos \theta$, $\theta$ is the angle between $\vec{k}$ and $\vec{B}$. The corresponding group velocity for the Alfven wave is $ v_g = \pm \frac{B_0 cos \theta}{ \sqrt{B_0^2 + \epsilon_0 + P_0 }} = \pm v_A cos\theta$, it is clear that the Alfven waves are unaffected from the metric perturbation. Interestingly, the gravitational sector shows explicit dependence on the magnetic field; the dispersion relation for gravitational modes are :
	\begin{equation}
		\label{eq:GWd}
		\omega = \pm  \sqrt{k^2 + 8 B_0^2 G \pi - 2 \Lambda +16  G \pi P_0 }.
	\end{equation}  
	Corresponding group velocity is $v_g = \pm \frac{k}{\sqrt{k^2 + 8 B_0^2 G \pi - 2 \Lambda + 16  G \pi P_0 }}$. As it is clear form the expression for $v_g$ that it may become purely imaginary if the denominator becomes negative, which corresponds to the cutoff $k_{c}=\pm\sqrt{-8 B_0^2 G \pi +2 \Lambda -16  G \pi P_0}$. Here  we 
	observe that instability arises for $k \leq k_c$ and the magnetic field plays an important role.				
	Also, It seems that zones with varying magnetic fields gives rise to dispersive gravitational waves. 
	
	The different hydrodynamical modes in the small $k$ limits are the following:
	For small k limit we have :\\
	$ \omega  = \begin{cases} \frac{i}{\tau_c} + \mathcal{O}(k^2), \\
		\pm \sqrt{2} \sqrt{4 B_0^2 \pi G - \Lambda + 8 \pi G P_0 - 4 B_0^2 \pi G \tan^2{\theta}}  + \mathcal{O} (k^2),\\
		\pm c_s k, \text{for } \theta =0, \\
		\pm v_A k , \text{for } \theta =0 ,\\
		\pm v_M k , \text{for } \theta = \frac{\pi}{4}.\end{cases}$
	where the expression for the magneto-sonic wave $v_M$ is cumbersome and hence is not given here (see Eq.\eqref{eq:v_M} in appendix \eqref{app:causality}), we also note that 
	$v_M$ diverges for $\theta=\pi/2$. Here we have found additional non-hydro modes arising due to gravity; these gravitational 
	modes are also unstable when $tan^2\theta\geq \pi \left(1+\frac{2P_{0}}{B_0^2}\right)-\frac{\Lambda}{4B_0^2G}$.
	
	In the large $k$ limit we have the following expression for $v_g$:
	\begin{widetext}
		
		\begin{eqnarray}\nonumber
			v_g^2 &=&  \left[\frac{ 2 v_A^2 \beta_{\Pi}}{4 \epsilon_0} + 
			\frac{\epsilon_0}{4 \epsilon_0 (\epsilon_0 + P_0 + B_0^2)} (B^2_0 (2  +2 c_s^2 cos^2 \theta) + 
			2 (1 + c_s^2) ( \beta_{\Pi}+ P_0))  
			\right] \\ \nonumber
			&& \pm \frac{1}{4 \epsilon_0 (\epsilon_0 + P_0 + B_0^2)} \sqrt{\left[  \left[  2 \beta_{\Pi} \left( B^2_0 cos^2 \theta + \epsilon_0 + P_0 \right) + \epsilon_0 \left(B^2_0 (2 + 2c_s^2 cos^2\theta)   + 2 (P_0 + c_s^2 P_0)\right) \right]^2  \right]} \\ \nonumber
			&& \pm \frac{1 }{4 \epsilon_0 (\epsilon_0 + P_0 + B_0^2)} \sqrt{\left[ 16 B^2_0 (-2 + cos 2 \theta) cos^2 \theta \epsilon_0 (\beta_{\Pi} + P_0) (B_0^2 + \epsilon_0 + P_0)  \right]}.
		\end{eqnarray}
	\end{widetext}
	Here in the large $k$ limit we have found the fast and slow magneto-sonic modes with causality bound as a function of $\theta$ and $B_0$. For consistency check, we set $\theta =0$ which gives 
	$ v_g^2  = \begin{cases} v_A^2 ,\text{ for slow magneto-sonic}, \\ \nonumber
		c_s^2 + \frac{\beta_{\Pi}}{\epsilon_0} .\text{ for fast magneto-sonic}.
	\end{cases}$. These expressions are identical to the ones found in.\cite{Biswas:2020rps}.
	Clearly, for $\theta=0$, the causality holds if $v_A^2 \leq 1$ and $\beta_{\Pi} \leq (1-c_s^2) \epsilon_0$. Similarly for $\theta =\frac{\pi}{2}$; $ v_g^2  = \begin{cases} 0 ,\text{ for slow magneto-sonic}, \\ \nonumber
		\frac{B_0^2 + (1+ c_s^2) (\beta_{\Pi} + P_0)}{B^2_0 + \epsilon_0 + P_0}, \text{for fast magneto-sonic}.
	\end{cases}$ \\ In this case, the causality requires $c_s^2 \leq \frac{\epsilon_0 - \beta_{\Pi}}{P_0 + \beta_{\Pi}}$.

	\begin{figure}
		\centering
		\includegraphics[width=0.48\textwidth]{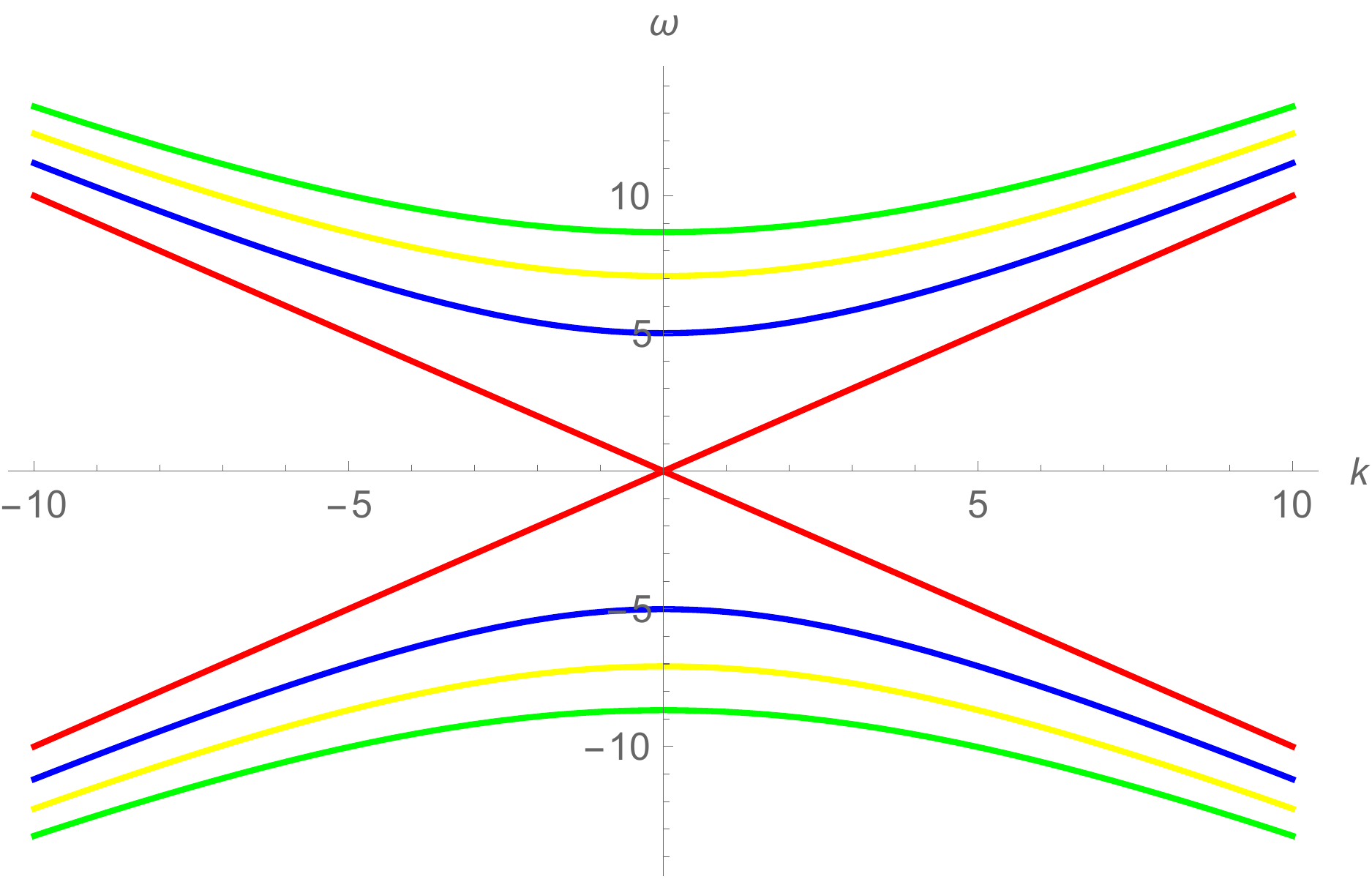}
		\includegraphics[width=0.48\textwidth]{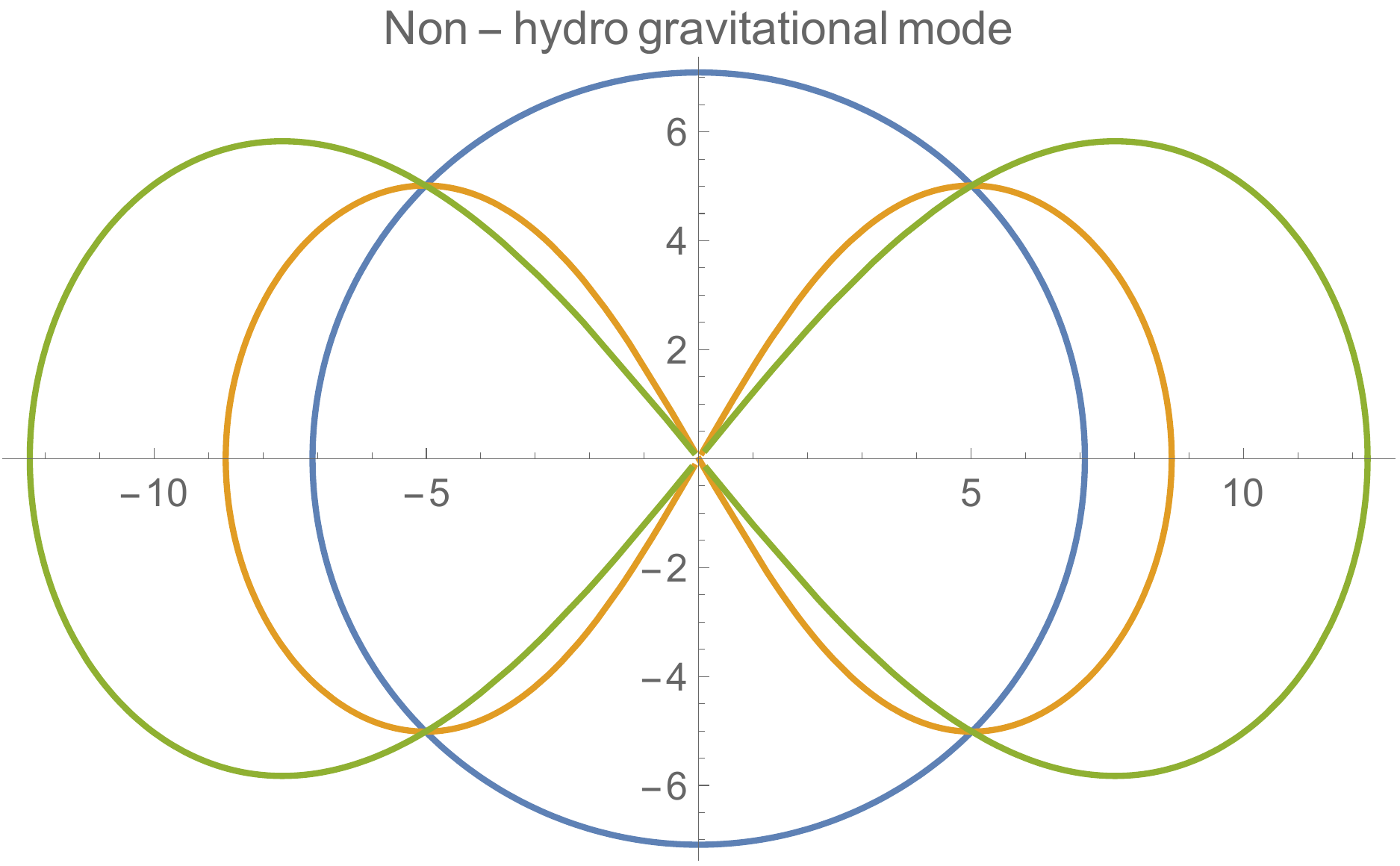}
		\caption{(Color online) Top panel: Gravitational modes $\omega^{\pm}$ vs $k$  for different pressure values (0,0.5,1.0) (in unit of $m_{\pi}^4$) for $B_0 = 1$ (in unit of $m_{\pi}^2$) , G=1 , $\Lambda =0 $, the one passing through the origin corresponds to ($B_0 =P_0 =0$).
			Bottom panel: Polar plot of the non-hydro modes due to gravity for different values of magnetic field (in unit of $m_{\pi}^2$)  for $G=1, \Lambda =0 , P_0 = 1 $.}
		\label{Fig:GW1}
		
	\end{figure}
	
	Top panel of Fig.\eqref{Fig:GW1} shows  the dispersion relation for the gravitational modes Eq.\eqref{eq:GWd} for different values of pressure and magnetic field. The gravitational waves travel on the light cone for $B_0 = P_0 =0$. But for non-zero $B_{0}$ and/or $P_{0}$ non-hydrodynamical modes arise as can be seen from Fig.\eqref{Fig:GW1} (top panel), the blue, yellow, and green lines correspond to $B_{0}=1$, $P_{0}=0-1.0$ in steps of 0.5 respectively. The bottom panel of Fig.\eqref{Fig:GW1}  is the polar plot of $\sqrt{2} \sqrt{4 B_0^2 \pi G - \Lambda + 8 \pi G P_0 - 4 B_0^2 \pi G \tan^2{\theta}}$ (the non-hydro modes due to gravity) as a function of $\theta$. As can be seen from the figure that in the small $k$ limit,  for $B_0$=0,  $\omega$ remains real $\forall$ $\theta$. Whereas, if we have finite magnetic field, then $\omega$ becomes complex for some ranges of $\theta$, and this angular dependence is a function of $B_{0}$; for example the yellow and green lines correspond to $B_{0}=1,2$ (in unit of $m_{\pi}^2$).

	
	\section{Summary and conclusion}
	\label{sec:summary}
	In this work, we formulate the first and second-order viscous magnetohydrodynamics for an arbitrary smooth (no-singularity) space-time geometry in the ideal-MHD limit (large magnetic Reynolds number). We found no explicit curvature dependence in the evolution of dissipative stresses as was found for the holographic calculation at finite temperature~\cite{Baier:2007ix} (without electromagnetic fields). This might not be surprising as it was mentioned in ref~\cite{Baier:2007ix} that for ordinary (i.e., without electromagnetic fields) relativistic hydrodynamics derived from kinetic theory using RTA approximation should not contain any explicit curvature dependence because the collision integral only allows symmetric tensors (for more details see ref~\cite{Baier:2007ix}). We further investigate wave propagation within the framework of linearised perturbation of fluid, field, and space-time metric. In this approach, for the case of a non-viscous fluid, we get the normal sound mode and the gravitational modes, which in the large $k$ limit gives rise to causal propagation of perturbations. We show that one of these gravitational modes leads to instability. After adding the bulk viscosity, we get the usual gravitational modes, which were found in the non-viscous case, with the causality and stability conditions remaining the same. 
	
	In the small $k$ limit, we found the usual non-hydro and the sound modes (arising due to bulk viscosity). All the hydrodynamic modes for this case are stable. On the other hand, for the large k limit, we found the causality condition to be $ \beta_{\Pi}\leq (1-c_s^2)\epsilon_0$. Lastly, the usual Alfven waves appear along with the magnetic field-dependent gravitational waves when we add the magnetic field. For this case, instability arises when $k$ becomes less than a cutoff value $k_{c}=\pm\sqrt{-8 B_0^2 G \pi +2 \Lambda -16 G \pi P_0}$. In the small $k$ limit, we found the usual non-hydro mode (due to the bulk viscosity), sound modes, and the Alfven modes. Interestingly, the magneto-sonic modes turned out to be dependent on the gravitational field and the angle between $\vec{k}$ and $\vec{B}$. We found additional non-hydro modes due to gravity which blows up at $\theta = \pi/2$. Whereas, in the large $k$ limit, only slow and fast magneto-sonic modes appear for which we have shown the corresponding causality conditions. 
	\footnote{While finishing our work we came across a similar study by Noh et al \cite{Noh:2020hgc} where they study the gravito-magnetic instabilities of a static homogeneous medium with magnetic field in the contexts of MHD with post-Newtonian (PN) corrections, and, special relativistic (SR) MHD with weak gravity.}
	In future, we plan to investigate the causality using real space analysis which takes into account non-linear effects. 
	
	\begin{acknowledgments}
		We would like to acknowledge Rajesh Biswas for his valuable inputs.
		AP acknowledges the CSIR-HRDG financial support. VR acknowledge support from the DAE, Govt. of India and financial support from the DST Inspire faculty research grant (IFA-16-PH-167), India.
	\end{acknowledgments}
	
	\begin{widetext}
		
		\appendix
		\section{Useful definitions}\label{app:definition}
		In this appendix we give details of a few definitions used in the main text. The relativistic thermodynamic integrals are defined as: 
		
		\begin{eqnarray}\nonumber
			J^{(m)\pm}_{nq}&=&\frac{1}{(2q+1)!!}\int \frac{\mathcal{G}_s d^3p}{(2\pi)^3 p_0}\sqrt{-g} (g_{\mu\nu}u^{\mu}p^{\nu})^{n-2q-m}\left(\Delta_{\alpha\beta}p^{\alpha}p^{\beta}\right)^q \left(f_0\tilde{f}_0\pm \bar{f}_0 \tilde{\bar{f}}_0\right),
		\end{eqnarray}
		here $\mathcal{G}_s$ is the spin degeneracy, where $n , q$ are integers with $n\ge 1$ and $q \ge 0$.
		The $n$-th moments integral for the distribution function is defined as:
		\begin{eqnarray}
			I_{\mu_{1}\mu_{2}\cdots\mu_{n}}^{(m)\pm}=\int \frac{\mathcal{G}_s d^3p}{(2\pi)^3 p_0 \left(u\cdot p\right)^{m }}\sqrt{-g} p_{\mu_{1}}p_{\mu_{2}}\cdots p_{\mu_{n}} \left(f_{0}\pm \bar{f_0}\right),
		\end{eqnarray}
		which can be docomposed as:
		\begin{eqnarray}
			\nonumber
			I_{\mu_{1}\mu_{2}\dots\mu_{n}}^{(m)\pm}=I_{n0}^{(m)\pm}u_{\mu_{1}}\cdots u_{\mu_{n}}+I_{n1}^{(m)\pm}
			\left(\Delta_{\mu_{1}\mu_{2}} u_{\mu_{3}} \cdots u_{\mu_{n}}+ \text{perm.}\right)+ \cdots \\ \label{eq:Iexpansion}
			\cdots + I_{nq}^{(m)\pm}\left(\Delta_{\mu_{1}\mu_{2}}\Delta_{\mu_{3}\mu_{4}}\cdots \Delta_{\mu_{n-1}\mu_{n}}
			+\text{perm.} \right).
		\end{eqnarray}
		where $n\geq 2q$.

		We also note that $J$
		is connected to the I integral via the following relation ,
		
		\begin{equation}\label{relJI}
			J^{(0)\pm}_{nq}=\frac{1}{\beta}\left[-I^{(0)\pm}_{n-1,q-1}+(n-2q)I^{(0)\pm}_{n-1,q}\right].
		\end{equation}
		
		Equilibrium number density , energy density  and pressure  are given as $I_{10}^{(0)-}$,  $I_{20}^{(0)+}$, $-I_{21}^{(0)+}$ respectively. Next we discuss the explicit form of $\delta f^{(1)}$ given in Eq.\eqref{eq:deltaf1}. For that we need the following derivatives:	
		\begin{eqnarray}
			\label{eq:gradf0}
			\partial_{\mu}f_0&=&-f_0\tilde{f}_0\left(g_{\alpha\beta}u^{\alpha}p^{\beta}\partial_{\mu}\beta+\beta \partial_{\mu}\left(g_{\alpha\beta}u^{\alpha}p^{\beta}\right)-\partial_{\mu}\alpha\right),\\ 
			\label{eq:gradpf0}
			\frac{\partial f_0}{\partial p^{\mu}}&=&-f_0 \tilde{f}_0 \beta u_{\sigma}\delta^{\sigma}_{\mu}.
		\end{eqnarray}
		Here $\delta^{s}_{\mu}$ is the Kronecker-Delta.We have also used the following formulas in the main text  as and whenever required:
		
		\begin{eqnarray}\nonumber
			\partial_{\mu}(\sqrt{-g}p^{\mu}f)&=&\sqrt{-g}p^{\mu}\partial_{\mu}f+\sqrt{-g}p^{\mu}\Gamma^{\alpha}_{\alpha\mu}f,\\ \nonumber
			\partial_{\mu}\left(2 \theta(p^0) \delta(p^2-m^2)\right)&=&2 \theta(p^0)\delta^{'}(p^2-m^2) 2 p^{\alpha}p_{\beta} \Gamma^{\beta}_{\alpha\mu},\\ \nonumber
			2p_{\beta}\delta^{'}(p^2-m^2)&=&\frac{\partial}{\partial p^{\beta}} \delta(p^2-m^2).
		\end{eqnarray}

		
		\section{$\delta f$ calculation}\label{app:2}
		In this appendix we give the detail derivation of first and second order corrections to the single particle distribution functions used 
		in the main text. First, let us discuss $\delta f^{(1)}$; using  Eq.\eqref{eq:deltafn}), Eq.\eqref{eq:gradf0}, and Eq.\eqref{eq:gradpf0} we have the following expression for $\delta f^{(1)}$:
		\begin{equation} \label{app:deltaf1}
			\delta f^{(1)}=f_0 \tilde{f}_0\frac{\tau_c}{u.p}\left((u.p)p^{\mu}\partial_{\mu}\beta +\beta p^{\beta}p^{\mu}D_{\mu}u_{\beta}-p^{\mu}\partial_{\mu}\alpha\right) +f_0 \tilde{f}_0\frac{\tau_c}{u.p}{\beta qF^{\mu\nu}p_{\nu}\delta^{\sigma}_{\mu}u_{\sigma}}.
		\end{equation}
		The covariant derivative of $D_{\mu}u_{\beta}$ in the above expression can be decomposed in the following manner
		\begin{equation}
			\label{app:covU}
			D_{\mu}u_{\beta}=u_{\mu}\dot{u}_{\beta}+\nabla_{\mu}u_{\beta},
		\end{equation}
		where $\dot{u}_{\beta}$ is the proper acceleration of fluid element.
		In a similar manner we can calculate the second-order $\delta f$ from Eq.\eqref{eq:deltafn}), that has the following form:
		\begin{equation}
			\label{eq:delf2}
			\delta f^{(2)}=-\frac{\tau_c}{u.p}\left(p^{\alpha}\partial_{\alpha} \delta f^{(1)}+qF^{\alpha\beta}p_{\beta}\frac{\partial \delta f^{(1)}}{\partial p^{\alpha}}-\Gamma^{\alpha}_{\nu\rho}p^{\nu}p^{\rho}\frac{\partial \delta f^{(1)}}{\partial p^{\alpha}}\right).
		\end{equation}
		Using Eq.\eqref{app:deltaf1}, Eq.\eqref{app:covU} in Eq.\eqref{eq:delf2} we have:
		\begin{eqnarray}\nonumber
			\delta f^{(2)} &=&  -\frac{\tau_c }{u.p}p^{\alpha}
			\left(\tau_c (\partial_{\mu}\beta)  p^{\mu} \left[-f_0 \tilde{f}_0\left(\tilde{f}_0-rf_0\right)\left((u.p) \partial_{\alpha}\beta+\beta p^{\beta}\left(D_{\alpha}u_{\beta} \right)-\partial_{\alpha}\alpha\right)\right]\right)\\ \nonumber
			&& -\frac{\tau_c^2 }{(u.p)^2}p^{\alpha}(\partial_{\mu}\alpha)  p^{\mu}f_0 \tilde{f}_0\left(\left[\left(\tilde{f}_0-rf_0\right) \left((u.p) \partial_{\alpha}\beta+\beta p^{\beta}\left(D_{\alpha}u_{\beta} \right)-\partial_{\alpha}\alpha\right)\right]+\frac{1}{(u.p)} p^{\beta}\left(D_{\alpha}u_{\beta} \right) \right)\\ \nonumber
			&&-\frac{\tau_c }{u.p}p^{\alpha}f_0 \tilde{f}_0 \left( p^{\mu} D_{\alpha}\left(\tau_c \partial_{\mu}\beta\right) +\tau_c \beta q E^{\nu}p_{\nu} \left[\frac{\left(\tilde{f}_0-rf_0\right)\left((u.p) \partial_{\alpha}\beta+\beta p^{\beta}\left(D_{\alpha}u_{\beta} \right)-\partial_{\alpha}\alpha\right) }{u.p}  \right] \right) \\ \nonumber
			&&+ \frac{\tau_c^2 \beta }{(u.p)^2}p^{\alpha}f_0 \tilde{f}_0 \left( -q E^{\nu}p_{\nu}\frac{1}{(u.p)} p^{\beta}\left(D_{\alpha}u_{\beta} \right)+ (D_{\mu}u_{\beta})p^{\beta}p^{\mu}\left[\left(\tilde{f}_0-rf_0\right) \left((u.p) \partial_{\alpha}\beta+\beta p^{\beta}\left(D_{\alpha}u_{\beta} \right)-\partial_{\alpha}\alpha\right) \right]\right)\\ \nonumber
			&&+\frac{\tau_c }{(u.p)^2}p^{\alpha}p^{\mu}f_0 \tilde{f}_0   \left(\beta \tau_c \left(D_{\mu} u_{\beta}\right)p^{\beta}  \frac{1}{(u.p)} p^{\nu}\left(D_{\alpha}u_{\nu}\right) + D_{\alpha}\left(\tau_c \partial_{\mu}\alpha\right) + qD_{\alpha} \left[\tau_c \beta E_{\mu}\right] \right)\\ \nonumber
			&&-\frac{\tau_c f_0 \tilde{f}_0 p^{\alpha}p^{\beta}p^{\mu}}{(u.p)^2} \left[\left(D_{\mu}u_{\beta}\right)\partial_{\alpha}\left(\beta \tau_c \right)+\beta \tau_c D_{\alpha}\left(D_{\mu}u_{\beta}\right)\right] \\ \nonumber
			&&-\frac{\tau_c^2}{(u.p)}q F^{\alpha\sigma} p_{\sigma}\left( \left[f_0 \tilde{f}_0 \partial_{\alpha}\beta-p^{\mu}(\partial_{\mu}\beta) f_0 \tilde{f}_0 \beta u_{\alpha}\left(\tilde{f}_0-rf_0 \right)\right]+\beta  \left[\frac{f_0 \tilde{f}_{0}}{u.p}\left(p^{\mu} D_{\alpha} u_{\mu} + p^{\beta} D_{\alpha} u_{\beta}\right)
			\right]\right)\\ \nonumber
			&&+\frac{\beta \tau_c^2}{(u.p)}q F^{\alpha\sigma} p_{\sigma}p^{\beta}p^{\mu} \left(\sigma_{\mu\beta}+\omega_{\mu\beta}+\frac{\Delta_{\mu\beta}\theta}{3}+u_{\mu}\dot{u}_{\beta}\right)\left[\frac{f_0 \tilde{f}_{0}}{(u.p)^2}u_{\alpha}+f_0 \tilde{f}_0 \beta u_{\alpha}\frac{\left(\tilde{f}_0-rf_0 \right)}{u.p}\right]\\ \nonumber
			&&+\frac{\tau_c^2}{(u.p)}q F^{\alpha\sigma} p_{\sigma}\left(\partial_{\mu}\alpha \left[\frac{f_0 \tilde{f}_{0}}{u.p}\delta^{\mu}_{\alpha}- p^{\mu}\left[\frac{f_0 \tilde{f}_{0}}{(u.p)^2}u_{\alpha}+f_0 \tilde{f}_0 \beta u_{\alpha}\frac{\left(\tilde{f}_0-rf_0 \right)}{u.p}\right] \right]+\beta q E_{\alpha}\left[\frac{f_0 \tilde{f}_{0}}{u.p} \right]\right)\\ \nonumber
			&&-\frac{ \beta q^2\tau_c^2}{(u.p)} F^{\alpha\sigma} p_{\sigma}E^{\nu}p_{\nu}\left[\frac{f_0 \tilde{f}_{0}}{(u.p)^2}u_{\alpha}+f_0 \tilde{f}_0 \beta u_{\alpha}\frac{\left(\tilde{f}_0-rf_0 \right)}{u.p}\right].\\ \nonumber
		\end{eqnarray}
		
		\subsection{Shear viscosity }
		We decompose the second-order expression for the shear stress  for the sake of convenient as:
		\begin{eqnarray}\nonumber
			\pi^{\lambda\chi }_{(2)}&=& I + II + III ,\\ \nonumber
		\end{eqnarray}
		where :
		\begin{eqnarray}
			I&=& -\frac{\Delta^{\lambda\chi}_{\alpha\beta}}{3}\tau_c \int \frac{d^4 p}{(2\pi)^3} \sqrt{-g} 2\Theta(p^0) \delta(p^2-m^2)  \frac{p^{\alpha}p^{\beta}p^{\kappa}}{(u.p)} \partial_{\kappa} \delta f^{(1)}, \\ \nonumber
			II &=& -\frac{\Delta^{\lambda\chi}_{\alpha\beta}}{3} \tau_c q F^{\mu}_{\quad \nu }\int \frac{d^4 p}{(2\pi)^3} \sqrt{-g} 2\Theta(p^0) \delta(p^2-m^2)  \frac{p^{\alpha}p^{\beta} p^{\nu}}{(u.p)} \frac{\partial \delta f^{(1)}}{\partial p^{\mu}},\\ \nonumber
			III &=&  \frac{\Delta^{\lambda\chi}_{\alpha\beta}}{3}\tau_c \Gamma^{\mu}_{\nu\rho} \int \frac{d^4 p}{(2\pi)^3} \sqrt{-g} 2\Theta(p^0) \delta(p^2-m^2) \frac{p^{\alpha}p^{\beta}p^\nu p^\rho}{(u.p)} \frac{\partial \delta f^{(1)}}{\partial p^{\mu}}.   
		\end{eqnarray}
		Let us evaluate each term one-by-one, for convenience let us define $\mathcal{I}^{\lambda\chi}_{\alpha\beta}\equiv  \frac{\Delta^{\lambda\chi}_{\alpha\beta}}{3}\tau_c \int \frac{d^4 p}{(2\pi)^3} \sqrt{-g} 2\Theta(p^0) \delta(p^2-m^2)$, then we have 
		\begin{eqnarray}\nonumber
			I&=&-  \mathcal{I}^{\lambda\chi}_{\alpha\beta} \frac{p^{\alpha}p^{\beta}p^{\kappa}}{(u.p)}\left( \tau_c \partial_{\mu}\beta  p^{\mu} \left[-f_0 \tilde{f}_0\left(\tilde{f}_0-rf_0\right)\left((u.p) \partial_{\kappa}\beta+\beta p^{\beta}\left(D_{\kappa}u_{\beta}+\Gamma^{\nu}_{\kappa\beta}u_{\nu} \right)-\partial_{\kappa}\alpha\right)\right]\right)\\ \nonumber
			&&+ \mathcal{I}^{\lambda\chi}_{\alpha\beta}\left(\tau_c (\partial_{\mu}\alpha)  p^{\mu}\left[\frac{\left(\tilde{f}_0-rf_0\right) \partial_{\kappa}f_0}{u.p}-  \frac{\left(f_0 \tilde{f}_0\right)}{(u.p)^2} p^{\beta}\left(D_{\kappa}u_{\beta}+\Gamma^{\nu}_{\kappa\beta}u_{\nu} \right)\right]\right) \\ \nonumber
			&&-  \mathcal{I}^{\lambda\chi}_{\alpha\beta}\left(-\tau_c \beta q E^{\nu}p_{\nu} \left[\frac{\left(\tilde{f}_0-rf_0\right) \partial_{\kappa}f_0}{u.p}-  \frac{\left(f_0 \tilde{f}_0\right)}{(u.p)^2} p^{\beta}\left(D_{\kappa}u_{\beta}+\Gamma^{\nu}_{\kappa\beta}u_{\nu} \right)\right]+f_0 \tilde{f}_0 p^{\mu} \partial_{\kappa}\left(\tau_c \partial_{\mu}\beta\right) \right)\\ \nonumber
			&&-  \mathcal{I}^{\lambda\chi}_{\alpha\beta}\left(\beta \tau_c (D_{\mu}u_{\beta})p^{\beta}p^{\mu}\left[\frac{\left(\tilde{f}_0-rf_0\right) \partial_{\kappa}f_0}{u.p}-  \frac{\left(f_0 \tilde{f}_0\right)}{(u.p)^2} p^{\beta}\left(D_{\kappa}u_{\beta}+\Gamma^{\nu}_{\kappa\beta}u_{\nu} \right)\right]\right)\\ \nonumber 
			&&+ \mathcal{I}^{\lambda\chi}_{\alpha\beta}\left(\left(\frac{f_0 \tilde{f}_0 p^{\mu}}{u.p}\right)\partial_{\kappa}\left(\tau_c \partial_{\mu}\alpha\right)+q\left(f_0 \tilde{f}_0 \frac{p_{\nu}}{u.p}\right)\left[ (\tau_c \beta)\partial_{\kappa} E^{\nu}+E^{\nu}\partial_{\kappa}(\tau_c \beta)\right]\right)\\ \nonumber
			&&-  \mathcal{I}^{\lambda\chi}_{\alpha\beta}\left(\left(\frac{f_0 \tilde{f}_0 p^{\beta}p^{\mu}}{u.p}\right) \left[\left(D_{\mu}u_{\beta}\right)\partial_{\kappa}\left(\beta \tau_c \right)+\beta \tau_c \partial_{\kappa}\left(D_{\mu}u_{\beta}\right)\right]\right),
		\end{eqnarray}

		\begin{eqnarray}\nonumber
			II &=& -  \mathcal{I}^{\lambda\chi}_{\alpha\beta} q  F^{\mu}_{\quad\nu} \frac{p^{\alpha}p^{\beta}p^{\nu}}{(u.p)} \tau_c  f_0 \tilde{f}_0 \partial_{\mu}\beta \\ \nonumber
			&&-  \mathcal{I}^{\lambda\chi}_{\alpha\beta} q  F^{\mu}_{\quad\nu}\left(\left[\tau_c p^{\kappa}(\partial_{\kappa}\beta) f_0 \tilde{f}_0 \beta u_{\mu}\left(-\tilde{f}_0+rf_0 \right)\right]+\beta \tau_c \left[\frac{f_0 \tilde{f}_{0}}{u.p}p^{\kappa}\left(p^{\mu}D_{\mu} u_{\alpha} + p^{\kappa} D_{\alpha} u_{\kappa}\right)\right]\right)\\ \nonumber
			&&-  \mathcal{I}^{\lambda\chi}_{\alpha\beta} q  F^{\mu}_{\quad\nu}\left(\beta \tau_c \left(\sigma_{\kappa\beta}+\omega_{\kappa\beta}+\frac{\Delta_{\kappa\beta}\theta}{3}+u_{\kappa}\dot{u}_{\beta}\right)p^{\beta}p^{\kappa}\left[-\frac{f_0 \tilde{f}_{0}}{(u.p)^2}u_{\mu}+f_0 \tilde{f}_0 \beta u_{\mu}\frac{\left(-\tilde{f}_0+rf_0 \right)}{u.p}\right]\right)\\ \nonumber
			&&+  \mathcal{I}^{\lambda\chi}_{\alpha\beta} q  F^{\mu}_{\quad\nu}\left(\tau_c \partial_{\kappa}\alpha \left[\frac{f_0 \tilde{f}_{0}}{u.p}\delta^{\kappa}_{\mu}+ p^{\kappa}\left[-\frac{f_0 \tilde{f}_{0}}{(u.p)^2}u_{\mu}+f_0 \tilde{f}_0 \beta u_{\mu}\frac{\left(-\tilde{f}_0+rf_0 \right)}{u.p}\right] \right]+\tau_c \beta q E_{\mu}\left[\frac{f_0 \tilde{f}_{0}}{u.p} \right]\right)\\ \nonumber
			&&+  \mathcal{I}^{\lambda\chi}_{\alpha\beta} q  F^{\mu}_{\quad\nu}\left(\tau_c \beta q E^{\nu}p_{\nu}\left[-\frac{f_0 \tilde{f}_{0}}{(u.p)^2}u_{\mu}+f_0 \tilde{f}_0 \beta u_{\mu}\frac{\left(-\tilde{f}_0+rf_0 \right)}{u.p}\right]\right), \\ \nonumber
		\end{eqnarray}

		\begin{eqnarray}\nonumber
			III &=& \mathcal{I}^{\lambda\chi}_{\alpha\beta}  \Gamma^{\mu}_{\nu\rho}  \frac{p^{\alpha} p^{\beta} p^{\nu} p^{\rho}}{(u.p)}\left(\tau_c  \left[f_0 \tilde{f}_0 \partial_{\mu}\beta\right]\right)\\ \nonumber
			&&+\mathcal{I}^{\lambda\chi}_{\alpha\beta}  \Gamma^{\mu}_{\nu\rho} \left(\left[\tau_c p^{\kappa}(\partial_{\kappa}\beta) f_0 \tilde{f}_0 \beta u_{\mu}\left(-\tilde{f}_0+rf_0 \right)\right]+\beta \tau_c \left[\frac{f_0 \tilde{f}_{0}}{u.p}p^{\kappa}\left(p^{\mu}D_{\mu} u_{\alpha} + p^{\kappa} D_{\alpha} u_{\kappa}\right)\right]\right)\\ \nonumber
			&&+\mathcal{I}^{\lambda\chi}_{\alpha\beta}  \Gamma^{\mu}_{\nu\rho}\left(\beta \tau_c \left(\sigma_{\kappa\beta}+\omega_{\kappa\beta}+\frac{\Delta_{\kappa\beta}\theta}{3}+u_{\kappa}\dot{u}_{\beta}\right)p^{\beta}p^{\kappa}\left[-\frac{f_0 \tilde{f}_{0}}{(u.p)^2}u_{\mu}+f_0 \tilde{f}_0 \beta u_{\mu}\frac{\left(-\tilde{f}_0+rf_0 \right)}{u.p}\right]\right)\\ \nonumber
			&&+\mathcal{I}^{\lambda\chi}_{\alpha\beta}  \Gamma^{\mu}_{\nu\rho}\left(-\tau_c \partial_{\kappa}\alpha \left[\frac{f_0 \tilde{f}_{0}}{u.p}\delta^{\kappa}_{\mu}+ p^{\kappa}\left[-\frac{f_0 \tilde{f}_{0}}{(u.p)^2}u_{\mu}+f_0 \tilde{f}_0 \beta u_{\mu}\frac{\left(-\tilde{f}_0+rf_0 \right)}{u.p}\right] \right]-\tau_c \beta q E_{\mu}\left[\frac{f_0 \tilde{f}_{0}}{u.p} \right]\right)\\ \nonumber
			&&+\mathcal{I}^{\lambda\chi}_{\alpha\beta}  \Gamma^{\mu}_{\nu\rho}\left(-\tau_c \beta q E^{\nu}p_{\nu}\left[-\frac{f_0 \tilde{f}_{0}}{(u.p)^2}u_{\mu}+f_0 \tilde{f}_0 \beta u_{\mu}\frac{\left(-\tilde{f}_0+rf_0 \right)}{u.p}\right]\right). \\ \nonumber
		\end{eqnarray}
		Combining the above expressions for I,II,III, and Eq.\eqref{eq:shear}  we get Eq.\eqref{eq:2evolutionexp}.
		\subsection{Bulk viscosity}
		We follow exactly same procedure to calculate $\Pi_{(2)}$ i.e., 
		\begin{eqnarray}\nonumber
			\Pi_{(2)}&=& I + II + III,
		\end{eqnarray}
		where 
		\begin{eqnarray}\nonumber
			I&=& \frac{\Delta_{\alpha\beta}}{3}\tau_c \int \frac{d^4 p}{(2\pi)^3} \sqrt{-g} 2\Theta(p^0) \delta(p^2-m^2)  \frac{p^{\alpha}p^{\beta}p^{\kappa}}{(u.p)} \partial_{\kappa} \delta f^{(1)}, \\ \nonumber
			II &=& \frac{\Delta_{\alpha\beta}}{3} \tau_c q F^{\mu}_{\quad \nu }\int \frac{d^4 p}{(2\pi)^3} \sqrt{-g} 2\Theta(p^0) \delta(p^2-m^2)  \frac{p^{\alpha}p^{\beta} p^{\nu}}{(u.p)} \frac{\partial \delta f^{(1)}}{\partial p^{\mu}},\\ \nonumber
			III &=& - \frac{\Delta_{\alpha\beta}}{3}\tau_c \Gamma^{\mu}_{\nu\rho} \int \frac{d^4 p}{(2\pi)^3} \sqrt{-g} 2\Theta(p^0) \delta(p^2-m^2) \frac{p^{\alpha}p^{\beta}p^\nu p^\rho}{(u.p)} \frac{\partial \delta f^{(1)}}{\partial p^{\mu}}.
		\end{eqnarray}
		
		For brevity let us define
		\begin{eqnarray}\nonumber
			\mathcal{I}_{\alpha\beta}&=& \frac{\Delta_{\alpha\beta}}{3}\tau_c \int \frac{d^4 p}{(2\pi)^3} \sqrt{-g} 2\Theta(p^0) \delta(p^2-m^2).
		\end{eqnarray}
		Now putting $\delta f^{(1)}$ and $\mathcal{I}_{\alpha\beta}$ in I,II, and III we get :

		\begin{eqnarray}\nonumber
			I&=& \mathcal{I}_{\alpha\beta} \frac{p^{\alpha}p^{\beta}p^{\kappa}}{(u.p)} \left( \tau_c (\partial_{\mu}\beta)  p^{\mu} \left[-f_0 \tilde{f}_0\left(\tilde{f}_0-rf_0\right)\left((u.p) \partial_{\kappa}\beta+\beta p^{\beta}\left(D_{\kappa}u_{\beta}+\Gamma^{\nu}_{\kappa\beta}u_{\nu} \right)-\partial_{\kappa}\alpha\right)\right]\right)\\ \nonumber
			&&-\mathcal{I}_{\alpha\beta} \frac{p^{\alpha}p^{\beta}p^{\kappa}}{(u.p)} \tau_c (\partial_{\mu}\alpha)  p^{\mu}\left[\frac{\left(\tilde{f}_0-rf_0\right) \partial_{\kappa}f_0}{u.p}-  \frac{\left(f_0 \tilde{f}_0\right)}{(u.p)^2} p^{\beta}\left(D_{\kappa}u_{\beta}+\Gamma^{\nu}_{\kappa\beta}u_{\nu} \right)\right] \\ \nonumber
			&&- \mathcal{I}_{\alpha\beta}\frac{p^{\alpha}p^{\beta}p^{\kappa}}{(u.p)}\left(\tau_c \beta q E^{\nu}p_{\nu} \left[\frac{\left(\tilde{f}_0-rf_0\right) \partial_{\kappa}f_0}{u.p}-  \frac{\left(f_0 \tilde{f}_0\right)}{(u.p)^2} p^{\beta}\left(D_{\kappa}u_{\beta}+\Gamma^{\nu}_{\kappa\beta}u_{\nu} \right)\right]-f_0 \tilde{f}_0 p^{\mu} \partial_{\kappa}\left(\tau_c \partial_{\mu}\beta\right) \right)\\ \nonumber
			&&+ \mathcal{I}_{\alpha\beta}\frac{p^{\alpha}p^{\beta}p^{\kappa}}{(u.p)}\left(\beta \tau_c (D_{\mu}u_{\beta})p^{\beta}p^{\mu}\left[\frac{\left(\tilde{f}_0-rf_0\right) \partial_{\kappa}f_0}{u.p}-  \frac{\left(f_0 \tilde{f}_0\right)}{(u.p)^2} p^{\beta}\left(D_{\kappa}u_{\beta}+\Gamma^{\nu}_{\kappa\beta}u_{\nu} \right)\right]\right)\\ \nonumber 
			&&- \mathcal{I}_{\alpha\beta}\frac{p^{\alpha}p^{\beta}p^{\kappa}}{(u.p)}\left(\left(\frac{f_0 \tilde{f}_0 p^{\mu}}{u.p}\right)\partial_{\kappa}\left(\tau_c \partial_{\mu}\alpha\right)+ q\left(f_0 \tilde{f}_0 \frac{p_{\nu}}{u.p}\right)\left[ (\tau_c \beta)\partial_{\kappa} E^{\nu}+E^{\nu}\partial_{\kappa}(\tau_c \beta)\right]\right)\\ \nonumber
			&&+ \mathcal{I}_{\alpha\beta}\frac{p^{\alpha}p^{\beta}p^{\kappa}}{(u.p)}\left(\left(\frac{f_0 \tilde{f}_0 p^{\beta}p^{\mu}}{u.p}\right) \left[\left(D_{\mu}u_{\beta}\right)\partial_{\kappa}\left(\beta \tau_c \right)+\beta \tau_c \partial_{\kappa}\left(D_{\mu}u_{\beta}\right)\right]\right),
		\end{eqnarray}
		
		\begin{eqnarray}\nonumber
			II &=&   q  F^{\mu}_{\quad\nu}  \mathcal{I}_{\alpha\beta} \frac{p^{\alpha}p^{\beta}p^{\nu}}{(u.p)}\tau_c f_0 \tilde{f}_0 \partial_{\mu}\beta\\ \nonumber
			&&+  q  F^{\mu}_{\quad\nu}  \mathcal{I}_{\alpha\beta} \frac{p^{\alpha}p^{\beta}p^{\nu}}{(u.p)}\left(\left[\tau_c p^{\kappa}(\partial_{\kappa}\beta) f_0 \tilde{f}_0 \beta u_{\mu}\left(-\tilde{f}_0+rf_0 \right)\right]+\beta \tau_c \left[\frac{f_0 \tilde{f}_{0}}{u.p}\left(p^{\mu}D_{\mu} u_{\alpha} + p^{\kappa} D_{\alpha} u_{\kappa}\right)\right]\right)\\ \nonumber
			&&+ q  F^{\mu}_{\quad\nu}  \mathcal{I}_{\alpha\beta} \frac{p^{\alpha}p^{\beta}p^{\nu}}{(u.p)}\left(\beta \tau_c \left(\sigma_{\kappa\beta}+\omega_{\kappa\beta}+\frac{\Delta_{\kappa\beta}\theta}{3}+u_{\kappa}\dot{u}_{\beta}\right)p^{\beta}p^{\kappa}\left[-\frac{f_0 \tilde{f}_{0}}{(u.p)^2}u_{\mu}+f_0 \tilde{f}_0 \beta u_{\mu}\frac{\left(-\tilde{f}_0+rf_0 \right)}{u.p}\right]\right)\\ \nonumber
			&&-  q  F^{\mu}_{\quad\nu}  \mathcal{I}_{\alpha\beta} \frac{p^{\alpha}p^{\beta}p^{\nu}}{(u.p)}\left(\tau_c \partial_{\kappa}\alpha \left[\frac{f_0 \tilde{f}_{0}}{u.p}\delta^{\kappa}_{\mu}+ p^{\kappa}\left[-\frac{f_0 \tilde{f}_{0}}{(u.p)^2}u_{\mu}+f_0 \tilde{f}_0 \beta u_{\mu}\frac{\left(-\tilde{f}_0+rf_0 \right)}{u.p}\right] \right]+\tau_c \beta q E_{\mu}\left[\frac{f_0 \tilde{f}_{0}}{u.p} \right]\right)\\ \nonumber
			&&-  q  F^{\mu}_{\quad\nu}  \mathcal{I}_{\alpha\beta} \frac{p^{\alpha}p^{\beta}p^{\nu}}{(u.p)}\left(\tau_c \beta q E^{\nu}p_{\nu}\left[-\frac{f_0 \tilde{f}_{0}}{(u.p)^2}u_{\mu}+f_0 \tilde{f}_0 \beta u_{\mu}\frac{\left(-\tilde{f}_0+rf_0 \right)}{u.p}\right]\right), \\ \nonumber
		\end{eqnarray}

		\begin{eqnarray}\nonumber
			III &=& - \Gamma^{\mu}_{\nu\rho} \mathcal{I}_{\alpha\beta}\frac{p^{\alpha} p^{\beta} p^{\nu} p^{\rho}}{(u.p)}\tau_c  f_0 \tilde{f}_0 \partial_{\mu}\beta\\ \nonumber
			&&- \Gamma^{\mu}_{\nu\rho} \mathcal{I}_{\alpha\beta}\frac{p^{\alpha} p^{\beta} p^{\nu} p^{\rho}}{(u.p)}\left(\left[\tau_c p^{\kappa}(\partial_{\kappa}\beta) f_0 \tilde{f}_0 \beta u_{\mu}\left(-\tilde{f}_0+rf_0 \right)\right]+\beta \tau_c \left[\frac{f_0 \tilde{f}_{0}}{u.p}p^{\kappa}\left(p^{\mu}D_{\mu} u_{\alpha} + p^{\kappa} D_{\alpha} u_{\kappa}\right)\right]\right)\\ \nonumber
			&&- \Gamma^{\mu}_{\nu\rho} \mathcal{I}_{\alpha\beta}\frac{p^{\alpha} p^{\beta} p^{\nu} p^{\rho}}{(u.p)}\left(\beta \tau_c \left(\sigma_{\kappa\beta}+\omega_{\kappa\beta}+\frac{\Delta_{\kappa\beta}\theta}{3}+u_{\kappa}\dot{u}_{\beta}\right)p^{\beta}p^{\kappa}\left[-\frac{f_0 \tilde{f}_{0}}{(u.p)^2}u_{\mu}+f_0 \tilde{f}_0 \beta u_{\mu}\frac{\left(-\tilde{f}_0+rf_0 \right)}{u.p}\right]\right)\\ \nonumber
			&&+ \Gamma^{\mu}_{\nu\rho} \mathcal{I}_{\alpha\beta}\frac{p^{\alpha} p^{\beta} p^{\nu} p^{\rho}}{(u.p)}\left(\tau_c \partial_{\kappa}\alpha \left[\frac{f_0 \tilde{f}_{0}}{u.p}\delta^{\kappa}_{\mu}+ p^{\kappa}\left[-\frac{f_0 \tilde{f}_{0}}{(u.p)^2}u_{\mu}+f_0 \tilde{f}_0 \beta u_{\mu}\frac{\left(-\tilde{f}_0+rf_0 \right)}{u.p}\right] \right]+\tau_c \beta q E_{\mu}\left[\frac{f_0 \tilde{f}_{0}}{u.p} \right]\right)\\ \nonumber
			&&+ \Gamma^{\mu}_{\nu\rho} \mathcal{I}_{\alpha\beta}\frac{p^{\alpha} p^{\beta} p^{\nu} p^{\rho}}{(u.p)}\left(\tau_c \beta q E^{\nu}p_{\nu}\left[-\frac{f_0 \tilde{f}_{0}}{(u.p)^2}u_{\mu}+f_0 \tilde{f}_0 \beta u_{\mu}\frac{\left(-\tilde{f}_0+rf_0 \right)}{u.p}\right]\right). \\ \nonumber
		\end{eqnarray}
		Using these above expressions for I, II,  III, and Eq.\eqref{eq:1bulk} we have Eq.\eqref{eq:2bulkevolutionexp}.
		\subsection{ Particle Diffusion}
		The calculation of the diffusion four-current  is exactly same:
		\begin{eqnarray}\nonumber
			V^{\mu}_{(2)}&=&I+II+III,
		\end{eqnarray}
		where	
		\begin{eqnarray}\nonumber
			I&=&-\tau_c \Delta^{\mu}_{\beta}\int dp  \frac{p^{\beta}p^{\alpha}}{u.p} \partial_{\alpha}  \delta f^{(1)}, \\ \nonumber
			II&=&-\tau_c q  F^{\alpha}_{\quad\rho}\Delta^{\mu}_{\beta}\int dp \frac{p^{\beta} p^{\rho} }{u.p}\frac{\partial \delta f^{(1)}}{\partial p^{\alpha}}, \\ \nonumber
			III&=& \tau_c \Delta^{\mu}_{\beta} \Gamma^{\alpha}_{\nu\rho}\int dp \frac{p^{\beta} p^{\nu}p^{\rho}}{u.p} \frac{\partial \delta f^{(1)}}{\partial p^{\alpha}}.
		\end{eqnarray}
		For convenience we define  
		\begin{eqnarray}\nonumber
			\mathcal{I}^{\mu}_{\beta} &=&  \tau_c \Delta^{\mu}_{\beta}\int \frac{d^4 p}{(2\pi)^3} \sqrt{-g} 2\Theta(p^0) \delta(p^2-m^2).
		\end{eqnarray}
		\begin{eqnarray}\nonumber
			I&=&- \mathcal{I}^{\mu}_{\beta} \frac{p^{\alpha}p^{\beta}}{u.p} \tau_c (\partial_{\mu}\beta)  p^{\mu} \left[-f_0 \tilde{f}_0\left(\tilde{f}_0-rf_0\right)\left((u.p) \partial_{\alpha}\beta+\beta p^{\beta}\left(D_{\alpha}u_{\beta}+\Gamma^{\nu}_{\alpha\beta}u_{\nu} \right)-\partial_{\alpha}\alpha\right)\right]\\ \nonumber
			&&+ \mathcal{I}^{\mu}_{\beta} \frac{p^{\alpha}p^{\beta}}{u.p}\tau_c (\partial_{\mu}\alpha)  p^{\mu}\left[\frac{\left(\tilde{f}_0-rf_0\right) \partial_{\alpha}f_0}{u.p}-  \frac{\left(f_0 \tilde{f}_0\right)}{(u.p)^2} p^{\beta}\left(D_{\alpha}u_{\beta}+\Gamma^{\nu}_{\alpha\beta}u_{\nu} \right)\right] \\ \nonumber
			&&+ \mathcal{I}^{\mu}_{\beta} \frac{p^{\alpha}p^{\beta}}{u.p}\left(\tau_c \beta q E^{\nu}p_{\nu} \left[\frac{\left(\tilde{f}_0-rf_0\right) \partial_{\alpha}f_0}{u.p}-  \frac{\left(f_0 \tilde{f}_0\right)}{(u.p)^2} p^{\beta}\left(D_{\alpha}u_{\beta}+\Gamma^{\nu}_{\alpha\beta}u_{\nu} \right)\right]-f_0 \tilde{f}_0 p^{\mu} \partial_{\alpha}\left(\tau_c \partial_{\mu}\beta\right) \right) \\ \nonumber
			&&- \mathcal{I}^{\mu}_{\beta} \frac{p^{\alpha}p^{\beta}}{u.p}\beta \tau_c (D_{\mu}u_{\beta})p^{\beta}p^{\mu}\left[\frac{\left(\tilde{f}_0-rf_0\right) \partial_{\alpha}f_0}{u.p}-  \frac{\left(f_0 \tilde{f}_0\right)}{(u.p)^2} p^{\beta}\left(D_{\alpha}u_{\beta}+\Gamma^{\nu}_{\alpha\beta}u_{\nu} \right)\right]\\ \nonumber 
			&&+ \mathcal{I}^{\mu}_{\beta} \frac{p^{\alpha}p^{\beta}}{u.p}\left(\left(\frac{f_0 \tilde{f}_0 p^{\mu}}{u.p}\right)\partial_{\alpha}\left(\tau_c \partial_{\mu}\alpha\right)+ q\left(f_0 \tilde{f}_0 \frac{p_{\nu}}{u.p}\right)\left[ (\tau_c \beta)\partial_{\alpha} E^{\nu}+E^{\nu}\partial_{\alpha}(\tau_c \beta)\right] \right)\\ \nonumber
			&&- \mathcal{I}^{\mu}_{\beta} \frac{p^{\alpha}p^{\beta}}{u.p}\left(\frac{f_0 \tilde{f}_0 p^{\beta}p^{\mu}}{u.p}\right) \left[\left(D_{\mu}u_{\beta}\right)\partial_{\alpha}\left(\beta \tau_c \right)+\beta \tau_c \partial_{\alpha}\left(D_{\mu}u_{\beta}\right)\right],
		\end{eqnarray}

		\begin{eqnarray}\nonumber
			II&=&- q F^{\alpha}_{\quad\rho}\mathcal{I}^{\mu}_{\beta} \frac{p^{\beta}p^{\rho}}{u.p} \left(\tau_c  \left[f_0 \tilde{f}_0 \partial_{\alpha}\beta+p^{\mu}(\partial_{\mu}\beta) f_0 \tilde{f}_0 \beta u_{\alpha}\left(-\tilde{f}_0+rf_0 \right)\right]+\beta \tau_c \left[\frac{f_0 \tilde{f}_{0}}{u.p}\left(p^{\mu}D_{\mu} u_{\alpha} + p^{\kappa} D_{\alpha} u_{\kappa}\right)
			\right]\right)\\ \nonumber
			&&- q F^{\alpha}_{\quad\rho}\mathcal{I}^{\mu}_{\beta} \frac{p^{\beta}p^{\rho}}{u.p}\beta \tau_c \left(\sigma_{\mu\beta}+\omega_{\mu\beta}+\frac{\Delta_{\mu\beta}\theta}{3}+u_{\mu}\dot{u}_{\beta}\right)p^{\beta}p^{\mu}\left[-\frac{f_0 \tilde{f}_{0}}{(u.p)^2}u_{\alpha}+f_0 \tilde{f}_0 \beta u_{\alpha}\frac{\left(-\tilde{f}_0+rf_0 \right)}{u.p}\right]\\ \nonumber
			&&- q F^{\alpha}_{\quad\rho}\mathcal{I}^{\mu}_{\beta} \frac{p^{\beta}p^{\rho}}{u.p}\left(-\tau_c \partial_{\mu}\alpha \left[\frac{f_0 \tilde{f}_{0}}{u.p}\delta^{\mu}_{\alpha}+ p^{\mu}\left[-\frac{f_0 \tilde{f}_{0}}{(u.p)^2}u_{\alpha}+f_0 \tilde{f}_0 \beta u_{\alpha}\frac{\left(-\tilde{f}_0+rf_0 \right)}{u.p}\right] \right]-\tau_c \beta q E_{\alpha}\left[\frac{f_0 \tilde{f}_{0}}{u.p} \right]\right)\\ \nonumber
			&&+ q F^{\alpha}_{\quad\rho}\mathcal{I}^{\mu}_{\beta} \frac{p^{\beta}p^{\rho}}{u.p}\tau_c \beta q E^{\nu}p_{\nu}\left[-\frac{f_0 \tilde{f}_{0}}{(u.p)^2}u_{\alpha}+f_0 \tilde{f}_0 \beta u_{\alpha}\frac{\left(-\tilde{f}_0+rf_0 \right)}{u.p}\right],
		\end{eqnarray}

		\begin{eqnarray}\nonumber
			III&=&\Gamma^{\alpha}_{\nu\rho} \mathcal{I}^{\mu}_{\beta}\frac{p^{\beta}p^{\nu}p^{\rho}}{u.p}\left(\tau_c  \left[f_0 \tilde{f}_0 \partial_{\alpha}\beta+p^{\mu}(\partial_{\mu}\beta) f_0 \tilde{f}_0 \beta u_{\alpha}\left(-\tilde{f}_0+rf_0 \right)\right]+\beta \tau_c \left[\frac{f_0 \tilde{f}_{0}}{u.p}\left(p^{\mu}D_{\mu} u_{\alpha} + p^{\kappa} D_{\alpha} u_{\kappa}\right)
			\right] \right)\\ \nonumber
			&&+\Gamma^{\alpha}_{\nu\rho} \mathcal{I}^{\mu}_{\beta}\frac{p^{\beta}p^{\nu}p^{\rho}}{u.p}\left(\beta \tau_c \left(\sigma_{\mu\beta}+\omega_{\mu\beta}+\frac{\Delta_{\mu\beta}\theta}{3}+u_{\mu}\dot{u}_{\beta}\right)p^{\beta}p^{\mu}\left[-\frac{f_0 \tilde{f}_{0}}{(u.p)^2}u_{\alpha}+f_0 \tilde{f}_0 \beta u_{\alpha}\frac{\left(-\tilde{f}_0+rf_0 \right)}{u.p}\right]\right)\\ \nonumber
			&&-\Gamma^{\alpha}_{\nu\rho} \mathcal{I}^{\mu}_{\beta}\frac{p^{\beta}p^{\nu}p^{\rho}}{u.p}\left(\tau_c \partial_{\mu}\alpha \left[\frac{f_0 \tilde{f}_{0}}{u.p}\delta^{\mu}_{\alpha}+ p^{\mu}\left[-\frac{f_0 \tilde{f}_{0}}{(u.p)^2}u_{\alpha}+f_0 \tilde{f}_0 \beta u_{\alpha}\frac{\left(-\tilde{f}_0+rf_0 \right)}{u.p}\right] \right]+\tau_c \beta q E_{\alpha}\left[\frac{f_0 \tilde{f}_{0}}{u.p} \right]\right)\\ \nonumber
			&&-\Gamma^{\alpha}_{\nu\rho} \mathcal{I}^{\mu}_{\beta}\frac{p^{\beta}p^{\nu}p^{\rho}}{u.p}\tau_c \beta q E^{\nu}p_{\nu}\left[-\frac{f_0 \tilde{f}_{0}}{(u.p)^2}u_{\alpha}+f_0 \tilde{f}_0 \beta u_{\alpha}\frac{\left(-\tilde{f}_0+rf_0 \right)}{u.p}\right].
		\end{eqnarray}
		Adding I, II, III, and Eq.\eqref{eq:diffusion} we have the desired result Eq.\eqref{eq:fulldiffusion}.

		\section{Causality and stability}\label{app:causality}
		In this appendix we discuss details of causality and stability given in Sec.\eqref{sec:causality} in the main text.
		For later use, we require the following expression for the $\delta\tilde{\Gamma}^{\mu}_{\nu\alpha}$:
		\begin{eqnarray}\label{app:deltagamma}
			\delta \tilde{\Gamma}^{\mu}_{\nu\alpha}&=&\frac{1}{2} \left[  i \delta \tilde{g}^{\mu}_{\alpha} k_{\nu} + i\delta \tilde{g}^{\mu}_{\nu} k_{\alpha}-i \delta \tilde{g}_{\nu\alpha} k^{\mu} \right].
		\end{eqnarray}
		We consider three different scenarios and discuss them one-by-one.
		\subsection{Non-viscous fluid with metric perturbation} 
		The first case we consider deals with non-viscous fluid in presence of metric perturbation.
		Various equations pertaining to this are :
		
		\begin{eqnarray}
			i \omega \delta \tilde{\epsilon}    +   \left(\epsilon_0 + P_0  \right)   i k_x \delta \tilde{u}^{x}  + ik_z \left(\epsilon_0 + P_0   \right)  \delta \tilde{u}^{z}
			&=&0, \\ 
			i \omega \left(\epsilon_0 + P_0  \right) \delta \tilde{u}^{x}  
			+ i k_x c_s^2 \delta \tilde{\epsilon}   
			&=&0,  \\ 
			i \omega \left(\epsilon_0 + P_0  \right) \delta \tilde{u}^{y} 
			&=&0,  \\ 
			i \omega  \left(\epsilon_0 + P_0  \right)\delta \tilde{u}^{z} 
			+ i k_z  c_s^2 \delta \tilde{\epsilon} 
			&=&0, \\ 
			8\pi G c_s^2 \delta \tilde{\epsilon}  - \left(\Lambda  + \frac{1}{2}   \left( \omega^2 - k^2 \right) - 8 \pi G P_0   \right) \delta \tilde{g}^{xx}   &=&  0,\\ 
			-\left(\Lambda  + \frac{1}{2}   \left( \omega^2 - k^2 \right) -
			8 \pi G P_0   \right) \delta \tilde{g}^{xy}
			&=& 0.
		\end{eqnarray}
		These above set of equations are used in writing Matrix $\mathcal{A}$ in sec.\eqref{Ideal fluid with Gravity:} 
		\subsection{  Viscous fluid with metric perturbation} 
		Let us now turn to the second case where we consider a fluid with non-zero bulk viscosity in presence of metric perturbations.
		The various equations pertaining to this is :
		\begin{eqnarray}
			i \omega \left(\delta \tilde{\epsilon}- \delta \tilde{\Pi}   \right)      +   i \left(\epsilon_0 + P_0 \right) k_x \delta \tilde{u}^{x}   + ik_z \left(\epsilon_0 + P_0   \right)  \delta \tilde{u}^{z}
			&=&0,  \\ 
			i \omega   \left(\epsilon_0 + P_0 \right) \delta \tilde{u}^{x}    
			+ i k_x  \left( c_s^2 \delta \tilde{\epsilon} + \delta \tilde{\Pi}  \right)      
			&=&0,  \\ 
			i \omega   \left(\epsilon_0 + P_0  \right) \delta \tilde{u}^{y}      
			&=&0,  \\ 
			i \omega   \left(\epsilon_0 + P_0 \right) \delta \tilde{u}^{z}    
			+ i k_z  \left( c_s^2 \delta \tilde{\epsilon} + \delta \tilde{\Pi}  \right)  &=&0 , \\ 
			\delta \tilde{\Pi} \left(  \frac{1}{\tau_c} + i \omega \right) + \beta_{\Pi} ik_{x} \delta \tilde{u}^{x}  + \beta_{\Pi} ik_{z} \delta \tilde{u}^{z}  &=& 0,  \\ 
			8\pi G \left( c_s^2 \delta \tilde{\epsilon} + \delta \tilde{\Pi}  \right)- \left(\Lambda  + \frac{1}{2}   \left( \omega^2 - k^2 \right)  - 8 \pi G P_0 \right) \delta \tilde{g}^{xx}  &=& 0, \\ 
			- \left(\Lambda  + \frac{1}{2}   \left( \omega^2 - k^2 \right)  - 8 \pi G P_0 \right) \delta \tilde{g}^{xy}&=& 0. 
		\end{eqnarray}
		These above set of equations are used in writing Matrix $\mathcal{A}$ in sec.\eqref{ Bulk viscosity with Gravity:}.		
		
		\subsection{ Viscous fluid with metric perturbation and magnetic field} 
		The last case we consider is applicable for a viscous fluid (only bulk viscosity) within the ideal MHD framework with metric perturbation.	
		Here in addition to the usual fluid variables we also need to consider magnetic field as an additional parameter. The linearised
		equations take the following form:
		\begin{eqnarray}
			i \omega \left(\delta \tilde{\epsilon}- \delta \tilde{\Pi}   \right)       +   \left(\epsilon_0 + P_0  \right)    i k_x \delta \tilde{u}^{x}     
			+ ik_z \left(\epsilon_0 +P_0   \right)  \delta \tilde{u}^{z}
			&=&0, \\ \nonumber
			i \omega   \left(\epsilon_0 + P_0 + B^2_0 \right) \delta \tilde{u}^{x}    
			+ i k_x \left[  c_s^2 \delta \tilde{\epsilon} + \delta \tilde{\Pi} -\frac{k_x}{\omega} \delta \tilde{u}^x  B^2_0 \right]  &&\\ 
			-  ik_z B^2_0  \delta \tilde{b}^{x}    + ik_x \left(\delta \tilde{g}^{xx} - \frac{k_x^2 + k_z^2}{k_z^2} \delta \tilde{g}^{xx}  \right) B_0^2 -i k_x \delta \tilde{g}^{xx} B^2_0  
			&=&0, \\ 
			i \omega  \left(\epsilon_0 + P_0 + B^2_0 \right) \delta \tilde{u}^{y}     
			- ik_{z} B^2_0 \delta \tilde{b}^{y} -   i \frac{k_x}{k_z} \delta \tilde{g}^{xy} k_{z} B^2_0 
			&=&0,  \\ \nonumber
			i \omega  \left(\epsilon_0 + P_0 + B^2_0 \right) \delta \tilde{u}^{z}     + i k_z  \left( c_s^2 \delta \tilde{\epsilon} + \delta \tilde{\Pi}  \right)   - i\omega B^2_0 \delta \tilde{u}^{z} &&\\ 
			+  ik_z \left(   \frac{3 k_x}{\omega} \delta \tilde{u}^x -  \delta \tilde{g}^{xx} + \frac{k_x^2 + k_z^2}{ k_z^2} \delta \tilde{g}^{xx}   \right)  B^2_0  - ik_x B^2_0 \delta \tilde{b}^{x}   &=&0,  \\ 
			\delta \tilde{\Pi} \left(  \frac{1}{\tau_c} + i \omega \right) + i\beta_{\Pi} k_{x} \delta \tilde{u}^{x}  + i\beta_{\Pi} k_{z} \delta \tilde{u}^{z}  &=& 0,  \\ 
			ik_z B_0 \delta \tilde{u}^{x}  -  i\omega  B_0 \delta \tilde{b}^{x} 
			&=& 0,  \\ 
			i k_z B_0  \delta \tilde{u}^{y}  - i\omega B_0 \delta \tilde{b}^{y}   &=& 0,  \\ 
			-\Lambda \delta \tilde{g}^{xx} - \frac{1}{2}  \delta \tilde{g}^{xx} \left( \omega^2 - k^2 \right) + 8\pi G \left( c_s^2 \delta \tilde{\epsilon} + \delta \tilde{\Pi} -\frac{k_x}{\omega} \delta \tilde{u}^x B^2_0 - \frac{  \delta \tilde{g}^{zz}}{2}B^2_0 \right)
			+ 8 \pi G \left(P_0  + \frac{B^2_0}{2} \right) \delta \tilde{g}^{xx}   &=& 0,\\ 
			-\Lambda \delta \tilde{g}^{xy} - \frac{1}{2}  \delta \tilde{g}^{xy} \left( \omega^2 - k^2 \right) + 8 \pi G \left(P_0 + \frac{B^2_0}{2} \right) \delta \tilde{g}^{xy} &=& 0.
		\end{eqnarray}
		These above set of equations are used in writing Matrix $\mathcal{A}$ in sec.\eqref{subsec:BulkMagneticGravity}.	\\
		The full expression for the $v_M$ which was not included in the main text for $\theta= \pi/4$ is as follows :
		\begin{eqnarray}\nonumber
			v_M &=& \pm \frac{1}{4 \left( B^2_0 + \epsilon_0 + P_0 \right) \left( 8 \pi G P - \Lambda \right)} \left[  - B^2_0 (2 + c_s^2) \Lambda - 2 (1+ c_s^2) \Lambda P_0 + 8 \pi G \left( B^2_0 (2- c_s^2) + B^2_0 c_s^2 (3 \epsilon_0 + 2 P_0) + 2 (P_0^2 + c_s^2 P_0^2) \right) \right] \\ \label{eq:v_M}
			&&\pm \frac{1}{4 \left( B^2_0 + \epsilon_0 + P_0 \right)} \left[\pm \sqrt{-16 B^2_0 c_s^2 (B^2_0 + \epsilon_0 + P_0) (\Lambda - 8 \pi G P_0) \left( -10 B^2_0 \pi G + \Lambda - 8 \pi G P_0\right) + \mathcal{T}}  \right],
		\end{eqnarray}
		
		where $$\mathcal{T}= \left[ 8 B^4_0 (c_s^2 -2) \pi G + B^2_0 (2 + c_s^2) \Lambda  + 2 P_0 \left( -4 B^2_0 (3 + 2 c_s^2) \pi G + \Lambda + \Lambda c_s^2 - 8 c_s^2 (\epsilon_0 + P_0) \pi G \right) \right]^2.$$
	\end{widetext}

\end{document}